\documentclass[12pt,letter]{article}

\usepackage{graphicx}
\usepackage{epstopdf}
\usepackage[centertags]{amsmath}
\usepackage{amsthm}
\usepackage{amssymb}
\usepackage{amsfonts}
\usepackage{ccaption}
\usepackage[usenames]{color}
\usepackage{mathrsfs}
\usepackage[square,sort&compress,numbers,comma]{natbib}

\usepackage[
      colorlinks=true,
      linkcolor=blue,  
      urlcolor=blue,    
      filecolor=black,     
      citecolor=red,
      pdfstartview=FitV,
      pdftitle={},
        pdfauthor={Mukund Rangamani},
        pdfsubject={Hydrodynamics},
        pdfkeywords={AdS/CFT, hydro, fluid-gravity},
        pdfpagemode=None,
        bookmarksopen=true    
      ]{hyperref}
       
\usepackage{rotating}

\makeatletter
\@addtoreset{equation}{section}

\makeatletter
\renewcommand\section{\@startsection {section}{1}{\z@}%
                                   {-3.5ex \@plus -1ex \@minus -.2ex}
                                   {2.3ex \@plus.2ex}%
                                   {\normalfont\large\bfseries}}
\renewcommand\subsection{\@startsection{subsection}{2}{\z@}%
                                     {-3.25ex\@plus -1ex \@minus -.2ex}%
                                     {1.5ex \@plus .2ex}%
                                     {\normalfont\bfseries}}

  \captionnamefont{\bfseries}
  \captiontitlefont{\small\sffamily}
  \captiondelim{: }
  \hangcaption

\def\sec#1{\S\ref{#1}}
\def\fig#1{Fig.\,\ref{#1}}
\def\req#1{(\ref{#1})}
\def\App#1{Appendix \ref{#1}}

\def\ie{{\it i.e.}}

\def\ord#1{\CO\left(#1\right)}

\def\thus{\Longrightarrow}

\def\CA{{\cal A}}
\def\CB{{\cal B}}
\def\CC{{ \cal C }}
\def\CD{{ \cal D }}

\def\CF{{\cal F}}

\def\CH{{\cal H}}

\def\CL{{\cal L}}

\def\CN{{\cal N}}
\def\CO{{\cal O}}

\def\CQ{{\cal Q}}
\def\CR{{\cal R}}
\def\CS{{\cal S}}
\def\CT{{\cal T}}

\def\CV{{\cal V}}

\def\R{{\bf R}}
\def\Sp{{\bf S}}

\def\A5S5{{\rm AdS}_5 \times \S^5}

\def\l{\ell}

\def\p{\partial}

\def\ord#1{\CO\left( #1 \right)}

\def\bz{b^{(0)}}
\def\bo{b^{(1)}}

\def\bCV{{\mathfrak V}}

\def\bCT{{\mathfrak T}}

\def\WD{{\CD}}

\def\AdS#1{AdS$_{#1}$}
\def\SAdS#1{Schwarzschild-AdS$_{#1}$}
\def\Schr#1{{\rm Schr}$_{#1}$}
\def\scri{\mathscr I}
\def\bCG{\mathfrak G}

\def\bcv{\mathfrak v}
\def\bcg{\mathfrak g}
\def\bCT{\mathfrak T}
\def\vel{v}
\def\vsq{v^2}
\def\u{u}
\def\DD{{\mathscr D}}
\title{{\bf Gravity \& Hydrodynamics:\\ Lectures on the fluid-gravity correspondence}}

\author{ Mukund Rangamani\footnote{mukund.rangamani@durham.ac.uk} \\ \\
\small{\emph{Centre for Particle Theory \& Department of
Mathematical Sciences}}, \\[-1.5mm]
\small{\emph{Science Laboratories, South Road, Durham DH1 3LE, United Kingdom}}
\\[1.5mm]
\small{\emph{Kavli Institute for Theoretical Physics}},\\[-1.5mm]
\small{\emph{University of California, Santa Barbara, CA 93015, USA}}
}

\begin{document}

\setlength{\baselineskip}{16pt}
\begin{titlepage}
\maketitle
\begin{picture}(0,0)(0,0)
\put(350, 340){DCPT-09/33}
\put(350,325){NSF-KITP-09-65}
\end{picture}
\vspace{-36pt}

\begin{abstract}
We discuss recent developments in the hydrodynamic description of strongly coupled conformal field theories using the AdS/CFT correspondence. In particular, we review aspects of the fluid-gravity correspondence which provides a map between a class of inhomogeneous, dynamical, black hole solutions in asymptotically AdS spacetimes and arbitrary fluid flows in the strongly interacting boundary field theory. We explain  how the geometric duals to the fluid dynamics are constructed in a boundary derivative expansion and use the construction to extract the hydrodynamic transport coefficients.  In addition we also describe the recent developments extending the correspondence to incorporate matter fields and to non-relativistic systems. Based on lectures given at the CERN Winter School on Supergravity, Strings and Gauge Theories, Geneva, Switzerland (February 2009).
 \end{abstract}
\thispagestyle{empty}
\setcounter{page}{0}
\end{titlepage}

\renewcommand{\baselinestretch}{1}  
\tableofcontents

\section{Introduction}
\label{s:intro}

One of the important  questions in modern theoretical physics involves understanding the dynamics of strongly coupled quantum field theories. Not only is this of theoretical interest, but there is a large class of real world physical systems where conventional perturbation theory is a poor description of the actual physics. A case in point which is partly relevant to the current discussion is the fascinating state of  matter discovered in heavy ion collisions, the Quark Gluon Plasma (QGP),  which is known to behave as a nearly ideal fluid. 

An important tool in the theoretician's toolkit to address strong coupling dynamics is the AdS/CFT correspondence \citep{Maldacena:1997re,Gubser:1998bc,Witten:1998qj}, which provides us with a holographic reformulation of field theory dynamics in terms of classical gravitational dynamics (in a higher dimensional spacetime). In general the AdS/CFT correspondence relates two seemingly disparate systems --  it provides a deep inter-connection between an interacting quantum field theory on the one hand and string theory in a curved background on the other. The field theories of interest are typically large $N$ gauge theories (obeying large $N$ factorization in the planar limit) and the holographic dual is in terms of string theory in an asymptotically AdS background.  While at generic values in parameter space one is dealing with two intrinsically complicated theories, at corners of parameter/coupling space one or the other description simplifies. Clearly, one has a simple field theoretic description when the coupling parameter is taken to be small; perturbation theory becomes reliable. In this regime the holographic dual description is in terms of string propagating in a highly curved background. In the opposite limit when we dial the field theory coupling to be strong, we simplify the string theory into classical (super-)gravity. 

There is a way of thinking about the AdS/CFT correspondence that chimes well with intuition for large $N$ gauge theories which is worth bearing in mind. On general grounds one expects that the large $N$ limit of a quantum gauge theory behaves effectively classically i.e., quantum fluctuations are suppressed by $1/N$. Said differently, the full quantum dynamics in this planar limit should be encoded in terms of a classical gauge field configurations, the ``Master field'' which controls the dynamics in this regime. While we have no concrete candidate for this Master field in large $N$ QCD, for a wide class of supersymmetric gauge theories which arise as the world-volume theories on D-branes, the AdS/CFT correspondence identifies a candidate Master field -- this is just string theory (or classical gravity) in an asymptotically AdS spacetime.
In the strong coupling regime of the field theory the dynamics of single trace operators is captured completely by classical Einstein gravity coupled to other fields. 

In the course of these lectures we are going to be interested in a specific limit of the correspondence -- we wish in particular to simplify the dynamics of the field theory to that of an effective classical fluid dynamics. As we shall see there is a precise sense in which this can be done for any interacting quantum field theory, by focussing on near-equilibrium dynamics and restricting attention to long wavelength physics. Under the holographic map we are led to consider a particular class of gravitational solutions, which turn out to be dynamical black hole spacetimes. This limit of the AdS/CFT correspondence which provides a concrete relation between the physics of fluids and that of gravity is what we call the {\it fluid-gravity correspondence} \citep{Bhattacharyya:2007lq}. In the course of these lectures we will derive this correspondence and see its utility in various contexts. Before delving into the detail however it is worthwhile to pause and take stock of the reasons for why this is an interesting endeavour.

\paragraph{Fluid dynamics:} As a classical dynamical system fluid dynamics provides interesting theoretical challenges. It is well known that for non-relativistic incompressible viscous fluids described by the Navier-Stokes equations the issue of finding globally regular solutions remains a open challenge, see for e.g., \citep{Fefferman:2000wo}.  At the same time fluid dynamical evolution shows very intriguing physics such as turbulence whose detailed understanding is still lacking. Furthermore, the behaviour of energy cascade in turbulent flows and the corresponding inverse cascade in lower dimensions are intriguing phenomena that beg for a better explanation.  

A holographic mapping of the fluid dynamical system into classical gravitational dynamics could in principle help in unearthing some of these mysteries, at the very least by providing a new perspective on the problem. To be sure, much of the physics of turbulence and global regularity are of interest in the context of non-relativistic, incompressible Navier-Stokes equations, while a natural realization of hydrodynamics in the fluid-gravity correspondence leads to relativistic conformal fluids. This however is not a primary obstacle, for as we will discuss towards the end, generalizations of the fluid-gravity correspondence to relax these constraints already exist. An obvious fantasy would be to hope that one can formulate a holographic dual of turbulence, but this is beyond the scope of these lectures.

\paragraph{Gravitational solutions:} Over the past decade new remarkable stationary solutions to higher dimensional gravity have been discovered and have served to highlight the limitations of the folk-lore about gravitational dynamics built on our intuition for gravity in four dimensions (for a review see \citep{Emparan:2008eg}). An excellent illustration of this are the black hole uniqueness theorems which fail to generalize straightforwardly to higher dimensions. The problem of stability of the solutions in higher dimensions is also reveals some differences from the lower dimensional analogs. One encounters for instance Gregory-Laflamme instabilities for black strings and black branes \citep{Gregory:1993vy} and closely related instabilities for spinning black holes \citep{Emparan:2003sy}. Understanding the stability domain of a given classical solution is important to  get a clear picture of the classical phase space of higher dimensional solutions.  

Fluid dynamics provides an interesting window to understand the physics of black holes. The idea of applying hydrodynamic intuition  to black holes dates back to the works on the membrane paradigm \citep{Thorne:1986eu,Damour:2008ji}, wherein one modeled the black hole horizon by a membrane equipped with fluid like properties. More recently, analog models for black hole stability problem have been proposed whereby the Rayleigh-Plateau instabilities of liquid droplets was associated with Gregory-Laflamme instabilities of black holes \citep{Cardoso:2006ks}. However, in these applications the fluid dynamics is merely an analogy, a mnemonic to understand the qualitative physical details in the complex gravitational setting by invoking a simpler fluid model. 

The fluid-gravity correspondence however provides a real duality between the hydrodynamic description and the  gravitational dynamics. This in particular implies that one can draw a precise quantitative connection between the two and thus enables us to understand aspects of the phase structure of black holes solutions and their stability in terms of the fluid model.  More pertinently for our current discussion this holographic duality also allows us to systematically construct dynamical black hole solutions. As we will see in the course of the lectures every fluid flow in the boundary field theory will map to a black hole spacetime in the  bulk geometric description with a regular event horizon. In fact, the fluid-gravity correspondence will enable us to algorithmically construct black hole geometries given solutions to the fluid equations of motion.\footnote{In fact there are several results in the literature exploring the phase structure \citep{Lahiri:2007ae,Bhardwaj:2008if,Bhattacharya:2009gm} and stability \citep{Caldarelli:2008mv,Maeda:2008kj} of black holes using the dual fluid dynamics in the context of field theories compactified on spatial circles with supersymmetry breaking boundary conditions. These boundary conditions break conformal invariance and the field theories are confining in the infra-red.}

\paragraph{Relevance to real world physics:}  Theoretical understanding of the state of matter  produced in  heavy-ion collisions at RHIC (and perhaps soon at LHC), the QGP, requires knowledge of dynamics in strongly coupled QCD. Current understanding is that subsequent to the collision of the ions, the resulting constituents of the system rapidly thermalize and comes into local thermal equilibrium and thenceforth evolve according to hydrodynamics until the local temperature falls back below the deconfinement temperature and the QGP hadronizes.  The hydrodynamic regime is characterized by a set of transport coefficients; in particular, since much of the flow in the QGP is the shear driven elliptic flow, it is the shear viscosity of the plasma that is of most relevance. Ideally, one would like to be able to start from the microscopic description in terms of QCD and be able to compute these transport coefficients. However, with the QCD coupling constant remaining strong near the deconfinement temperature one needs to find a way to go beyond perturbation theory. The obvious choice, Lattice QCD is somewhat handicapped in this respect since it is ill-equipped to deal with real time physics and Lorentzian correlation functions.\footnote{For recent developments on lattice computations see \citep{Aarts:2008yw} and references therein.} 

The AdS/CFT correspondence provides a theoretical framework to understand some of the qualitative features of the hydrodynamics seen in QGP, by providing us with an efficient way to access strongly coupled physics in a class of superconformal field theories. While these field theories are {\em qualitatively} different from QCD in their vacuum, one might argue that at finite temperature some of these differences are perhaps mitigated. An interesting observation based on lattice simulations of QCD free energy, is that there appears to be a range of temperatures say between $2 \,T_c$ and $5\, T_c$ (recall that $T_c \sim 175$\,MeV for QCD) where the energy density as a function of temperature shows Stefan-Boltzmann scaling with a numerical pre-factor which is about $80\%$ of the free field value. This could be taken as prima-facie evidence for an effective description as a  strongly coupled CFT for it is very similar to the situation in $\CN=4$ Super-Yang Mills (SYM) theory, whose strong coupling free energy is exactly $3/4$ the free field value \citep{Gubser:1996de}.  However, other observables such as pressure deviate from the value predicted by conformal invariance, thereby weakening the analogy.  It  is therefore worth keeping in mind that the theories one is discussing are not quite QCD.

That said it is rather remarkable that the only class of strong coupling field theories for which we can compute hydrodynamic transport coefficients exhibit a remarkable quantitative agreement with those arising from numerical fits to RHIC data.  For instance the universal behaviour of the shear viscosity in hydrodynamic description of field theories with gravitational holographic duals  \citep{Kovtun:2004de} has already attracted attention and has impacted experimental analysis of RHIC data, see for example \citep{Shuryak:2003xe,Shuryak:2004cy,Shuryak:2006se,Shuryak:2008eq}.\footnote{For a recent account of the nearly ideal fluids in encountered nature, such as the QCP and cold atoms at unitarity, see \citep{Schaefer:2009dj}.}  In any event, independent of applications to heavy-ion collisions one can view the superconformal field theories as toy models; it is certainly quite remarkable that the holographic map allows us to explicitly determine the transport properties of a strongly coupled non-abelian plasma.

 \paragraph{Summary of the lectures:} In these lectures we use the AdS/CFT correspondence to study the effective description of strongly coupled conformal field theories at long wavelengths. On physical grounds it is reasonable that any interacting quantum field theory equilibrates locally at high enough energy densities,  
and so admits an effective description in terms of fluid 
dynamics. The variables of such a description are the local 
densities of all conserved charges together with the local 
fluid velocities. The equations of fluid dynamics are simply the equations 
of local conservation of the corresponding charge currents and energy-momentum tensor, supplemented by  constitutive relations that express these currents as functions of fluid mechanical variables. As fluid dynamics is a long wavelength effective theory, 
these constitutive relations are usually specified in a derivative expansion. 
At any given order, thermodynamics plus symmetries determine 
the form of this expansion up to a finite number of undetermined 
coefficients. These coefficients may then be obtained either from 
measurements or from microscopic computations.

The best understood examples of the AdS/CFT correspondence relate the 
strongly coupled dynamics of certain (super-)conformal field theories to the 
dynamics of gravitational systems in AdS spaces. 
In particular, we will demonstrate that Einstein's equations with a negative 
cosmological constant, supplemented with appropriate regularity 
restrictions and boundary conditions, reduce to the nonlinear equations 
of fluid dynamics in an appropriate regime of parameters. We provide 
a systematic framework to construct this universal nonlinear fluid dynamics, 
order by order in a boundary derivative expansion i.e., as an effective theory. 

There is a rather rich history of studying hydrodynamics of non-abelian plasmas using holographic methods provided by the AdS/CFT correspondence. The early work of \citep{Horowitz:1999jd} was the first to relate the process of thermalization in the field theory with the study of black hole quasi-normal modes. Subsequently, the seminal work of   Policastro, Son and Starinets \citep{Policastro:2001yc} began the investigation of  linearized fluid dynamics from linearized gravity in asymptotically AdS black hole backgrounds. The exploration of linearized hydrodynamics has been carried out in various different contexts over the years in \citep{Herzog:2002fn,Policastro:2002tn,Policastro:2002se,Son:2002sd,Herzog:2002pc,Herzog:2003ke,Kovtun:2003wp,Buchel:2003tz,Buchel:2004di,Buchel:2004qq,Kovtun:2004de,Kovtun:2005ev,Benincasa:2005iv,Maeda:2006by,Mas:2006dy,Saremi:2006ep,Son:2006em,Benincasa:2006fu,Myers:2008yi, Garousi:2008ai, Ghodsi:2009hg} 
and we refer the reader to the excellent review \citep{Son:2007vk} for other references on the subject. The fluid-gravity correspondence \citep{Bhattacharyya:2007lq} itself was motivated in part by these  studies and the attempts to construct the holographic dual of the so called  Bjorken flow \citep{Janik:2005zt,Janik:2006gp}\footnote{For further explorations of the spacetime geometry dual to Bjorken flow see  \citep{Nakamura:2006ih,Sin:2006pv, Janik:2006ft, Friess:2006kw, Kajantie:2006ya, Heller:2007qt,Kajantie:2007bn,Benincasa:2007tp,Natsuume:2007ty, Natsuume:2008iy} and \citep{Heller:2008fg} for a review. More recently, these class of geometries have been understood within the framework of the fluid-gravity correspondence in \citep{Heller:2008mb,Kinoshita:2008dq,Figueras:2009iu}. }  which is believed to be relevant to understanding the central region of heavy-ion collisions hydrodynamically \citep{Bjorken:1982qr}.   At the same time the investigations of the fluid dynamical regime of stationary black holes in asymptotically AdS spacetimes \citep{Bhattacharyya:2007vs} paved the way for  a clear understanding of the hydrodynamic regime in the gravitational context.

The fluid-gravity correspondence was originally discussed in \citep{Bhattacharyya:2007lq} in the context of gravitational duals of four dimensional superconformal field theories whose holographic dual is given by \AdS{5} $\times X_5$ where $X_5$ a Sasaki-Einstein manifold which determines the CFT. A special case is the $\CN =4$ SYM where $X_5 = \Sp^5$.  In \citep{Bhattacharyya:2008xc} the global aspects of the bulk geometry were discussed and a geometric construction of the fluid entropy current was given.  Subsequently, this discussion was generalized to other dimensions in  \citep{VanRaamsdonk:2008fp,Haack:2008cp, Bhattacharyya:2008mz}. In addition \citep{Bhattacharyya:2008ji} described how to include external forcing in the hydrodynamic description by placing the fluid on a curved manifold (and also explicitly included dilaton couplings). There is also a discussion of charged fluid dynamics  \citep{Erdmenger:2008rm,Banerjee:2008th,Hur:2008tq,Torabian:2009qk}, which in the context of $\CN =4$ SYM corresponds to looking at the grand-canonical ensemble with chemical potential for $U(1)_R$ charges, and in addition inclusion of magnetic and dyonic charges in \AdS{4} \citep{Hansen:2008tq,Caldarelli:2008ze}. While all these discussions are for conformal fluids whose duals are asymptotically AdS spacetimes, one can extend the discussion to non-conformal fluids living on Dp-brane world-volumes \citep{Kanitscheider:2009as,David:2009np} as well as explorations involving higher derivative terms in the gravitational description \citep{Dutta:2008gf}. Finally, the restriction of relativistic invariance can be relaxed to consider non-relativistic conformal fluids \citep{Rangamani:2008gi} as well as duals to incompressible Navier-Stokes flow \citep{Bhattacharyya:2008kq}. For a review of some of the developments on the subject see also \citep{Ambrosetti:2008mt}. We will mainly review the basic features of the correspondence in these lectures and will discuss some of the generalizations towards the end in \sec{s:generalz}. 

The essential physical points arising from the fluid-gravity correspondence  which we will describe in detail below can be summarized as follows:
\begin{itemize}
\item The gravitational derivation of the relativistic Navier-Stokes equations and its higher-order generalizations confirms the basic intuition that fluid dynamics is indeed the correct long-wavelength effective description of  strongly coupled field theory dynamics.
\item The geometries dual to fluid dynamics turn out to be black hole spacetimes with regular event horizons \citep{Bhattacharyya:2008xc}. This  indicates that the hydrodynamic regime is special and in particular can  be interpreted to indicate that the 
fluid dynamical stress tensors lead to regular gravity solutions respecting cosmic censorship.
\item The explicit construction of the fluid dynamical stress tensor leads to a precise determination of higher order transport coefficients for the dual field theory. The results we present for the transport coefficients for four dimensional superconformal field theories with holographic duals were also derived in \citep{Baier:2007ix}.
\end{itemize}

\paragraph{Outline of the lectures:} We will begin with a review of relativistic fluid dynamics in \sec{s:fluids} and then discuss various aspects of conformally invariant fluids in \sec{s:confflu}, reviewing in particular the extremely useful Weyl covariant formalism developed in \citep{Loganayagam:2008is} in \sec{s:weylcov}.  In \sec{s:gravity} we will review the basic scheme to construct gravitational solutions dual to fluid dynamics following \citep{Bhattacharyya:2007lq}. We then turn to a discussion of the physical properties of our solutions in \sec{s:grav2}. Finally in \sec{s:generalz} we will discuss various generalizations of the fluid-gravity correspondence and conclude with a discussion in \sec{s:discuss}.

\section{Elements of fluid dynamics}
\label{s:fluids}

Fluid dynamics is  the low energy effective description of any interacting quantum field theory, valid for fluctuations that are of sufficiently  long wavelength. This description is intrinsically statistical in nature, for it is the collective physics of a large number of microscopic constituents. Usually one thinks of the hydrodynamic description as the effective continuum model valid on macroscopic scales. For an excellent introduction to the basic ideas in fluid dynamics see \citep{Landau:1965pi} (see also \citep{Andersson:2007gf} for relativistic fluids). 

To understand the statistical origins of the fluid dynamical description, consider a quantum system in a grand canonical ensemble, where we prescribe the temperature and chemical potentials for various conserved charges. In global thermal equilibrium we calculate observables by computing correlation functions in the grand canonical density matrix. One can also perturb away from this global equilibrium scenario and allow the thermodynamic variables to fluctuate. For fluctuations whose wavelengths  are large compared to the scale set by the local energy density or temperature, one describes the system in terms of fluid dynamics. 

One can pictorially think of the situation as follows; sufficiently long-wavelength fluctuations variations are slow on the scale of the local energy density/temperature. Then about any given point in the system we expect to encounter a domain where the local temperature is roughly constant -- in this domain we can use the grand canonical ensemble to extract the physical characteristics of the field theory. Of course, different domains will described by different values for the intrinsic thermodynamic variables. Fluid dynamics describes how these different domains interact and exchange thermodynamic quantities. 

A more formal way to define the hydrodynamic regime is the following. In any interacting system there is an intrinsic length scale, the mean free path length $\ell_{\text{mfp}}$, which is the characteristic length scale of the interacting system. This is most familiar from the kinetic theory picture of gases, but applies equally well to fluids. In the kinetic theory context the $\ell_{\text{mfp}}$ simply characterizes the length scale for the free motion of the constituents between successive interactions. To achieve the hydrodynamic limit we are simply requiring that we examine the system at length scales which are large compared to the $\ell_{\text{mfp}}$, so that the microscopic inhomogeneities are sufficiently smeared out.

The dynamical content of the hydrodynamic equation is simply conservation of energy and other conserved global charges. These can be succinctly summarized by giving the stress tensor $T^{\mu\nu}$, which is a symmetric two tensor, and the charge currents $J^\mu_I$, where $I = \{1,2, \cdots\}$ indexes the set of conserved charges characterizing the system. The dynamical equations then are given as 
\begin{equation}
\nabla_\mu T^{\mu\nu} = 0 \  ,\qquad \nabla_\mu J^\mu_I = 0 \ .
\label{cons1}
\end{equation}	

To specify the system further we just need to find expressions for the stress tensor and charge currents.  Since the fluid dynamical description is only valid when the underlying microscopic QFT achieves local thermal equilibrium, we should be able to use the basic thermodynamic variables to characterize the system. Furthermore, we need to describe how the different domains of local thermally equilibrated fluid interact with each other. To understand how this can be achieved, let us focus on a fluid element which is characterized by  the local values of the thermodynamic variables. As the fluid element can exchange its characteristics with neighbouring fluid elements, we should associate to this fluid element a velocity field $u^\mu$, which describes the flow of thermodynamic quantities. It turns out that the thermodynamic variables together with the velocity field serve to characterize the fluid completely. 

Consider a QFT living on a $d$-spacetime dimensional background $\CB_d$ with (non-dynamical) metric $g_{\mu\nu}$. The coordinates on $\CB_d$ will be denoted henceforth as $x^\mu$. We can summarize the dynamical degrees of freedom characterizing the hydrodynamic description of this interacting QFT   as:
\begin{itemize}
\item Extrinsic quantities: Local energy density $\rho$ and charge densities $q_I$. 
\item Fluid velocity $u_\mu$ (normalized $g^{\mu \nu} \, u_\mu \, u_\nu = -1$).
\item Intrinsic quantities: Pressure $P$, temperature $T$,  and chemical potentials $\mu_I$ determined by equation of state.
\end{itemize}
All that remains is to express the stress tensor and charge currents in terms of these variables.

\subsection{Ideal fluids}
\label{s:idealf}

Let us focus first on the description of an ideal fluid which has no dissipation. Then by passing to a local rest frame, where we choose the velocity field to be aligned in the direction of energy flow, we can identify the components of the stress tensor as the energy density (temporal component longitudinal to the flow) and pressure (spatial components transverse to the velocity field). Similarly, in this local rest frame the components of the charge current are the charge density itself (along the velocity field). Putting this together, we learn that for an ideal fluid 
\begin{eqnarray}
\left(T^{\mu \nu}\right)_{\text{ideal}} &=& \rho\, u^\mu \, u^\nu + P\,\left(g^{\mu \nu} + u^\mu \, u^\nu\right)  \ ,  \nonumber \\
\left(J^\mu_I\right)_{\text{ideal}} &=& q_I \, u^\mu  .
\label{ideal1}
\end{eqnarray}	

Before proceeding to incorporate the physics of dissipation, let us pause to introduce some notation that will be useful in what follows. Since the $d$-velocity $u^\mu$ is oriented along the temporal direction, we can use it to decompose the  spacetime into spatial slices with induced metric
\begin{equation}\label{Pdef}
P^{\mu \nu} = g^{\mu \nu} + u^{\mu} \, u^{\nu}  .
\end{equation}
We can view $P^{\mu \nu}$ as a projector onto spatial directions; it is a symmetric positive definite tensor which satisfies the following identities:
\begin{equation}\label{Pprops}
P^{\mu \nu} \, u_{\mu} = 0 \ , \qquad
P^{\mu \rho} \, P_{\rho \nu} = P^{\mu}_{\ \nu} = P^{\mu \rho} \, g_{\rho \nu} \ , \qquad
P_{\mu}^{\ \mu} = d-1  .
\end{equation}
In terms of this projector we can express the ideal fluid stress tensor more simply as 
\begin{equation}
\left(T^{\mu \nu}\right)_{\text{ideal}}= \rho\, u^\mu \, u^\nu + P \, P^{\mu \nu}  . 
\label{ideal2}
\end{equation}	
While the decomposition of $T^{\mu\nu}$ into the temporal and spatial components is not very interesting at the level of ideal fluids, it will play a more important role when we include the effects of dissipation. 

In addition to the stress tensor and charge currents, there is another quantity which we would like to keep track of, viz., the entropy current. Pictorially the entropy current keeps track of how the local  entropy density varies in the fluid. For an ideal fluid given the entropy density $s(x)$ the entropy current takes the simple form
\begin{equation}
\left(J^\mu_S\right)_{\text{ideal}} = s\, u^\mu  . 
\label{entcideal}
\end{equation}	
Using the equations of motion \req{cons1}  for an ideal fluid \req{ideal1} and standard thermodynamic relations, it is easy to show that the entropy current is conserved 
\begin{equation}
\nabla_\mu \left(J^\mu_S\right)_{\text{ideal}}  =0 .
\label{entidealcons}
\end{equation}	
i.e., the fluid flow involves no production of entropy.

\subsection{Dissipative fluids}
\label{s:dissf}

Ideal fluids with stress tensor given by \req{ideal2} are an approximation;  they don't include any physics of dissipation. This is clearly seen by the conservation the entropy current. In general we expect that the flow of the fluid results in the creation of entropy consistent with the second law of thermodynamics. More pertinently, dissipation is necessary for a fluid dynamical system to equilibrate when perturbed away from  a given equilibrium configuration. Microscopically one can understand the dissipative effects as arising from the interaction terms in the underlying QFT. As a result the terms incorporating the effects of dissipation in the stress tensor and charge currents will depend on the coupling constants of the underlying QFT.

To model a hydrodynamical system incorporating the effects of dissipation we only need to add extra pieces to the stress tensor and charge currents. Let us denote the dissipative part of the stress tensor by $\Pi^{\mu\nu}$ as the corresponding part of the charge currents by $\Upsilon^\mu$:
\begin{eqnarray}
\left(T^{\mu \nu}\right)_{\text{dissipative}} &=& \rho\, u^\mu \, u^\nu + P\,\left(g^{\mu \nu} + u^\mu \, u^\nu\right)  + \Pi^{\mu\nu} \ ,  \nonumber \\
\left(J^\mu_I\right)_{\text{dissipative}} &=& q_I \, u^\mu + \Upsilon ^\mu .
\label{dissp1}
\end{eqnarray}	
To complete specification of the hydrodynamic system we should determine $\Pi^{\mu\nu}$ and $\Upsilon ^\mu$ in terms of the dynamical variables the velocity field, $u^\mu$, and the thermodynamic variables $\rho$, $P$, $q_I$, etc..

The traditional way to determine the dependence of the dissipative components of the stress tensor and charge currents is somewhat phenomenological. One employs the second law of thermodynamics (positive divergence of the entropy current), cf., \citep{Landau:1965pi,Andersson:2007gf}, to determine the set of allowed terms in the most general form of the constitutive equations. While this makes some sense for incorporating the leading non-linearities in the fluid description, at higher orders the procedure starts to be fraught with ambiguities as exemplified by the Israel-Stewart formalism \citep{Israel:1976tn,Israel:1979wp}. We will take a different viewpoint on this and in fact outline a procedure to unambiguously construct a non-linear stress tensor for hydrodynamics.

The procedure we have in mind is inspired by the manner one constructs effective field theories. When one writes down an effective field theory for a QFT,  at any given order one takes into account all possible terms that can appear in the effective lagrangian consistent with the underlying symmetry. These irrelevant operators are suppressed by power counting in momenta and appear with arbitrary coefficients in the effective lagrangian.  We have argued that hydrodynamics is best thought of as an effective classical description of any interacting field theory valid when the system achieves local thermal equilibrium. Hence following the logic of effective field theories we should allow in the hydrodynamic approximation all possible operators consistent with the symmetries. These irrelevant operators in the present case are nothing but derivatives of the velocity field and thermodynamic variables. Thus one can cleanly express the fluid dynamical stress tensor in a gradient expansion.\footnote{The power counting scheme we have in mind for fluid dynamical operators is the following: each spacetime derivative will count as dimension 1 for book keeping purposes. It is important to bear in mind that the actual scaling dimension of the operator (as discussed in \sec{s:confflu}) which follows from the transformation of the stress tensor under scale transformations is different.}

Before proceeding to enumerate the  various operators that can contribute to the stress tensor, let us constrain the dissipative components. First of all we have a choice of frame which is related to how we choose the fluid velocity. Secondly when we enumerate gradient terms at any given order we are free to use the lower order equations of motion to reduce the set of gradient terms. Let us discuss these two issues in turn and see how they serve to constrain the dissipative terms.

\paragraph{Defining the velocity field:} Since we are discussing relativistic fluids, the fluid velocity field needs to be defined with some care. The issue is simply that in relativistic dynamics one cannot distinguish between mass and energy in a clean fashion. Usually in non-relativistic fluids one would talk about heat flow between different fluid elements, but for a relativistic fluid heat flux necessarily leads to mass or momentum transfer and hence to energy flow. By picking the velocity field appropriately we can fix this ambiguity; the precise fashion in which we choose to do so is simply a matter of convention.

For ideal fluids we picked the velocity field such that in the local rest frame of a fluid element, the stress tensor components longitudinal to the velocity gave the local energy density in the fluid. One can in fact demand the same to be true when we incorporate the effects of dissipation. This leads to the so called {\em Landau frame} which is defined by demanding that the dissipative contributions be orthogonal to the velocity field i.e,.
\begin{equation}
\Pi^{\mu\nu} \, u_\mu = 0 \ , \qquad \Upsilon ^\mu\, u_\mu =0 \ .
\label{landauframe}
\end{equation}	
Formally, one can define the Landau frame as follows: the stress tensor which is a symmetric two-tensor on the background manifold $\CB_d$ has a single timelike eigenvector.  In the Landau frame we define the velocity field $u^\mu$ to be given by this eigenvector (unit normalized), so that the definition of the velocity field is tied to the energy-momentum transport in the system.  For charged fluids one has the possibility of working in the so called Eckart frame where the fluid velocity is tuned to charge transport -- here we define the velocity field to be the determined by the charge current. In what follows we will work exclusively in the Landau frame and hence apply \req{landauframe} to constrain the dissipative terms of the stress tensor and charge currents.

\paragraph{Enumerating independent operators:} The issue of enumerating the independent operators at any given order is completely analogous to the discussion in effective field theory. Suppose we are interested in generalizing the ideal fluid \req{ideal1} to first order in the gradient expansion. For the stress tensor we would write down all possible symmetric two tensors built out of the gradients of the velocity field and thermodynamic variables. However, the ideal fluid equations of motion \req{cons1} themselves are first order in derivatives, and therefore relate the derivatives of the thermodynamic potentials to the gradients of  the velocity field. Employing these relations we can simplify the expression for the first order stress tensor to be given in terms of derivatives of the velocity field alone. This process can clearly be iterated to higher orders. 

\paragraph{Decomposing velocity gradients:} We are interested in analyzing various combinations of derivatives of the velocity field.  First let us consider the decomposition of the velocity gradient $\nabla^{\nu}u^{\mu}$ into a part along the velocity, given by the {\it acceleration} $a^{\mu}$, and a transverse part.  The latter can in turn be decomposed into a trace, the {\it divergence} $\theta$, and the remaining traceless part with symmetric and antisymmetric components, respectively given by the {\it shear} $\sigma^{\mu \nu}$ and {\it vorticity} $\omega^{\mu \nu}$.
The decomposition can then be written as follows:
\begin{equation}\label{udec}
\nabla^{\nu}u^{\mu} 
= - a^{\mu} \, u^{\nu} 
+ \sigma^{\mu \nu} 
+ \omega^{\mu \nu} 
+ \frac{1}{d-1} \, \theta \, P^{\mu\nu} \ ,
\end{equation}
where the divergence, acceleration, shear, and vorticity, are defined as:\footnote{Note that we use standard symmetrization and anti-symmetrization conventions. For any tensor $F_{ab}$ we define the symmetric part $F_{(ab)} = \frac{1}{2}\, \left(F_{ab} + F_{ba}\right) $ and the anti-symmetric part $F_{[ab]} = \frac{1}{2}\left( F_{ab} - F_{ba}\right)$ respectively. We also use $\DD$ to indicate the velocity projected covariant derivative: $\DD \equiv u^\mu \, \nabla_\mu$.} 
\begin{equation}\label{udecdefs}
\begin{split}
\theta &= \nabla_{\mu}u^{\mu} = P^{\mu\nu} \, \nabla_{\mu}u_{\nu}  \\
a^{\mu}&= u^{\nu} \, \nabla_{\nu}u^{\mu} \equiv \DD u^{\mu} \\
\sigma^{\mu\nu} &= \nabla^{(\mu}u^{\nu)} + u^{(\mu} \, a^{\nu)}  - \frac{1}{d-1} \, \theta \, P^{\mu\nu}
=  P^{\mu \alpha} \, P^{\nu \beta} \, \nabla_{(\alpha} u_{\beta)}  - \frac{1}{d-1} \, \theta \, P^{\mu\nu} \\
\omega^{\nu\mu} &= \nabla^{[\mu}u^{\nu]} + u^{[\mu} \, a^{\nu]} 
= P^{\mu \alpha} \, P^{\nu \beta} \, \nabla_{[\alpha} u_{\beta]} \ .
\end{split}
\end{equation}
For future reference we note that we will also have occasion to use a the following notation to indicate symmetric traceless projections transverse to the velocity field. For any two tensor $\CT^{\mu\nu}$ we define:
\begin{equation}
\CT^{\langle\mu\nu\rangle} = P^{\mu\alpha} \, P^{\nu\beta}\, \CT_{(\alpha \beta)} - \frac{1}{d-1}\, P^{\mu\nu} \, P^{\alpha \beta}\, \CT_{\alpha\beta} \ .
\label{angleb}
\end{equation}	

Note that we can write the projectors a bit more compactly:
$ P^{\mu \alpha} \, P^{\nu \beta} \, \nabla_{(\alpha} \, u_{\beta)} = P^{\rho(\mu} \nabla_ \rho u^{\nu)}$ and $ P^{\mu \alpha} \, P^{\nu \beta} \, \nabla_{[\alpha} \, u_{\beta]} = P^{\rho[\mu} \nabla_ \rho u^{\nu]}$.
It is easy to verify all the previously asserted properties, 
in addition to $u_{\mu}\,a^{\mu} = 0$ and $P_{\mu\nu}\,a^{\mu} = a_{\nu}$:
\begin{equation}\label{udecprops}
\begin{split}
& \sigma^{\mu \nu} \, u_{\mu} = 0 \ , \qquad
\sigma^{\mu \rho} \, P_{\rho \nu} = \sigma^{\mu}_{\ \nu} \ , \qquad
\sigma_{\mu}^{\ \mu} = 0 \ , \\
& \omega ^{\mu \nu} \, u_{\mu} = 0 \ , \qquad
\omega ^{\mu \rho} \, P_{\rho \nu} = \omega ^{\mu}_{\ \nu} \ , \qquad
\omega_{\mu}^{\ \mu} = 0 \ .
\end{split}
\end{equation}

We are now in possession of sufficient amount of data to write down the dissipative part of the stress tensor to leading order in the derivative expansion. First of all let us notice that the zeroth order equations of motion i.e., those arising from the ideal fluid description relate the gradients of the energy density and pressure to those of the $u^\mu$. The quickest way to derive the required relation is to consider projections of the conservation equation $\nabla_\mu \left(T^{\mu\nu}\right)_{\text{ideal}} = 0$, along the velocity field and transverse to it, i.e.,
\begin{eqnarray}
u^\nu \, \nabla_\mu \left(T^{\mu\nu}\right)_{\text{ideal}} &=& 0 \;\; \thus\;\; 
(\rho + P)\, \nabla_\mu u^\mu + u^\mu \nabla_\mu \rho =0
\nonumber \\ 
\qquad P_{\nu \alpha} \,\nabla_\mu \left(T^{\mu\nu}\right)_{\text{ideal}} &=& 0 \;\; \thus \;\; P_\alpha^{\ \mu} \nabla_\mu P + (\rho+P)\, P_{\nu\alpha} \,u^\mu\,\nabla_\mu u^\nu =0 \ .
\label{zeroeom}
\end{eqnarray}	
 respectively. To characterize the stress-tensor at leading order in the gradient expansion our task is reduced to writing down symmetric two tensors that can be built solely from velocity gradients and we should furthermore account for the Landau frame condition. These conditions in fact isolate just two terms which can appear in the expression for $\Pi^{\mu \nu}$: 
\begin{equation}
\Pi^{\mu\nu}_{(1)} = -2\, \eta \, \sigma^{\mu \nu} - \zeta\, \theta\, P^{\mu\nu} \ ,
\label{pi1a}
\end{equation}	
where we have introduced two new parameters the {\it shear viscosity}, $\eta$, and the {\it bulk viscosity}, $\zeta$. 

Likewise for the charge current $\Upsilon^\mu$ we will obtain contributions which are first order in the derivatives of the thermodynamic variables $\rho$ and $q_I$ and also the velocity field (where now the acceleration contributes). We will choose this time to eliminate the acceleration term using zeroth order equation of motion 
\req{zeroeom} and express derivatives of energy density and charge density.
Usually these are the only contributions which are considered for the charge current. However, there is another potential contribution from a  pseudo-vector \citep{Erdmenger:2008rm, Banerjee:2008th}, 
\begin{equation}
\ell^\mu = \epsilon_{\alpha \beta \gamma}^{\hspace{5.5mm}\mu} \, u^\alpha\, \nabla^\beta u^\gamma ,
\label{lmudef}
\end{equation}	
provided we are willing to mix contributions with different parity structure at the first order. In fact, for fluid dynamics derived from the AdS/CFT correspondence for the $\CN =4$ Super Yang-Mills fluid, there is indeed a contribution of this sort to the charge current (in a specific ensemble).\footnote{This contribution can be traced to Chern-Simons couplings of the bulk gauge field in the gravitational description, c.f., \citep{Erdmenger:2008rm, Banerjee:2008th}.} Putting things together we find that we have a first order contribution to the charge current:
\begin{equation}
\Upsilon ^\mu_{(1)I} = -\widetilde{\varkappa}_{IJ} \, P^{\mu\nu} \, \nabla_{\nu} q_J - \widetilde{\gamma}_I\,  P^{\mu\nu} \, \nabla_{\nu} \rho - \mho_I\, \ell^\mu \ , 
\label{Gam1a}
\end{equation}	
where we again have new coefficients, $\widetilde{\varkappa}_{IJ}$ is the matrix of charge diffusion coefficients, $\widetilde{\gamma}_I$ indicates the contribution of the energy density to the charge current, and $\mho_I$ which are the pseudo-vector transport coefficients. For later purposes,  it will  be convenient to re-express the charge current in terms of the intensive parameters $\mu_I$ and $T$ (chemical potential and temperature), so that \req{Gam1a} can be recast equivalently as\footnote{One can relate the transport coefficients $\{\varkappa_{IJ},\gamma_I\}$  to the original coefficients $\{\widetilde{\varkappa}_{IJ}, \widetilde{\gamma}_I\}$ in terms of various susceptibilities. However, getting explicit expressions for these in general will require knowledge of the precise thermodynamic functions. We thank R. Loganayagam for useful discussion in this regard.}
\begin{equation}
\Upsilon ^\mu_{(1)I} = -\varkappa_{IJ} \, P^{\mu\nu} \, \nabla_{\nu} \left(\frac{\mu_J}{T}\right) - \mho_I\, \ell^\mu - \gamma_I\, P^{\mu\nu}\,\nabla_\nu T \ .
\label{Gam1a}
\end{equation}	

Assembling all the pieces together we claim that a generic charged fluid flow will satisfy the dynamical equations \req{cons1} with the stress tensor and charge currents at leading order in gradient expansion (dropping therefore higher derivative operators):
\begin{eqnarray}
T^{\mu \nu}&=& \rho\, u^\mu \, u^\nu + P\,\left(g^{\mu \nu} + u^\mu \, u^\nu\right)  -2\, \eta \, \sigma^{\mu \nu} - \zeta\, \theta\, P^{\mu\nu}\ ,  \nonumber \\
J^\mu_I &=& q_I \, u^\mu -\varkappa_{IJ} \, P^{\mu\nu} \, \nabla_{\nu} \left(\frac{\mu_J}{T}\right) -  \mho_I\, \ell^\mu - \gamma_I\, P^{\mu\nu}\,\nabla_\nu T \ .
\label{viscous1}
\end{eqnarray}	
Following conventional terminology we will refer to the fluid dynamical system specified  by \req{viscous1} as a viscous fluid or relativistic Navier-Stokes equations.\footnote{We will review the conventional non-relativistic equations later, see \sec{s:nonrel}.} 

Apart from the thermodynamic potentials, we have a set of transport coefficients $\eta$, $\zeta$, $\varkappa_{IJ}$, $\gamma_I$ and $\mho_I$ which need to be determined to completely specify the relativistic viscous fluid. If the QFT whose hydrodynamic description we seek is weakly coupled then we can in principle determine these coefficients in perturbation theory. This is per se not a trivial exercise, for a discussion of the computation of transport coefficients in perturbative gauge theory see 
\citep{Arnold:2000dr,Arnold:2003zc}. We will however be interested in deriving these coefficients for a strongly coupled quantum field theory using a dual holographic description.

Finally, let us turn to the other quantity of interest, viz., the entropy current. Including the effects of dissipation we learn that entropy is no longer conserved. Under fluid evolution we should have entropy production since the system is generically evolving from a non-equilibrium state to an equilibrium state. Entropy increase consistent with the second law can be rephrased as the statement that the entropy current should have non-negative divergence. For a relativistic viscous fluid we will therefore have an entropy current $J^\mu_S$ with 
\begin{equation}
\nabla_\mu J^\mu_S \ge 0 \ .
\label{seclaw}
\end{equation}	
It is possible to show that for uncharged fluids involving pure energy momentum transport one has a entropy current statisfying:
\begin{equation}
\nabla_\mu J^\mu_S  =\frac{2\,\eta}{T}\, \sigma_{\alpha\beta}\, \sigma^{\alpha\beta} \ ,
\label{divec1}
\end{equation}	
which is non-negative as long as the shear viscosity is positive.  We will return to a detailed discussion of the entropy current for conformal fluids in \sec{s:entcur}.

\subsection{Causality issues in relativistic viscous fluids}
\label{s:causal}

A natural question is whether the relativistic viscous fluids can be extended to higher orders.\footnote{We will refer the generalized system of fluid dynamical equations as non-linear viscous hydrodynamics, notwithstanding the fact that the relativistic Navier-Stokes equations themselves are non-linear.} To determine this we should just follow the conventional philosophy of effective field theory and try to write down the full set of operators at the desired order in gradient expansion. These higher order terms will be less and less important (they are irrelevant in the low energy effective theory) as we go to longer wavelengths and are thus somewhat inconsequential for the hydrodynamic evolution. The leading order terms which we incorporated in the viscous fluid \req{viscous1} are singled out because of their importance in realizing a channel for the fluid to relax back to equilibrium -- they are  the leading dissipative terms in the theory.  

However, in conventional relativistic fluid literature one usually encounters a different rationale having to do with causality issues and the initial value problem. Consider the viscous fluid system \req{viscous1}  with the dynamical equations \req{cons1}. This system of equations is first order in time derivatives,\footnote{The simplest way to see this is to work in the local inertial frame and realize that the gradient terms in \req{viscous1} only involve spatial derivatives owing to the projector $P^{\mu\nu}$.} and as a result the system of partial differential equations equations describing a relativistic viscous fluid is parabolic. 

In order to have a relativistic system with a well defined initial value problem the equations of motion are required to be hyperbolic. Only for hyperbolic partial differential equations is it possible to evolve initial data specified on a Cauchy surface. Usually lack of hyperbolicity leads to formation of singularities and manifests itself by acausal behaviour. A useful analogy to keep in mind, which is in fact central to the current discussion, is the heat equation, which being first order in time derivatives is parabolic. Solutions to the heat equation show diffusive behaviour which leads to acausal propagation of signals (when viewed with respect to the  light-cone of the underlying Lorentzian manifold $\CB_d$).

This issue has led several people to come up with so called ``causal relativistic hydrodynamics'', the most prominent among which is the phenomenologically motivated Israel-Muller-Stewart formalism \citep{Israel:1976tn,Israel:1979wp}.\footnote{For reviews on this subject we refer the reader to \citep{Andersson:2007gf,Muronga:2003ta}. In the context of heavy-ion collisions an excellent discussion complemtenting our presentation here can be found in \citep{Romatschke:2009im}. See also \citep{Natsuume:2007ty} for related discussion in the holographic context.} For sake of completeness we will briefly review the Israel-Stewart formalism and in particular will argue that the issue of causality violation in the hydrodynamic description is a red-herring.

The Israel-Stewart idea is to add higher order  terms to the relativistic viscous fluid system \req{viscous1}.  The terms that are added are such that the resulting entropy current satisfies the second law \req{seclaw}. However, it is worth noting that their phenomenological formalism doesn't fully account for all the irrelevant operators one can add at the corresponding order \citep{Baier:2007ix}. We will illustrate the formalism with a simple toy model which captures the basic idea behind their approach. 

Consider the derivation of the diffusion equation from the continuity equation, which is a conservation equation\footnote{For the sake of simplicity we will momentarily switch to a non-relativistic system; for the present discussion $\nabla$ is a spatial gradient.}
\begin{equation}
\frac{\partial \rho}{\partial t} + \nabla\cdot \vec{j} = 0 \ ,
\label{ctyeq}
\end{equation}	
and the phenomenological Fick's law
\begin{equation}
\vec{j} = -D \, \nabla \rho \ ,
\label{fick}
\end{equation}	
where $D$ is the diffusion constant. Eliminating the current $\vec{j}$ we find the diffusion equation
\begin{equation}
\frac{\p\rho}{\p t} - D \, \nabla^2 \rho = 0 \ .
\label{diffeq}
\end{equation}	

The equation  \req{diffeq} of course has a diffusive spectrum $\omega \sim - i\, \vec{k}\cdot \vec{k}$ and is a simple example of a parabolic PDE which doesn't have a well defined initial value problem. The problem is that any perturbation to the system instantaneously spreads out to arbitrary spatial scales -- essentially the fluctuations don't relax quickly enough. In the diffusion equation \req{diffeq}, due to absence of  quadratic term in time derivatives, the velocity of propagation of disturbances is infinite. Equivalently, in the absence of density fluctuations there is nothing forcing the current to relax to its stationary equilibrium value.  Ideally, the current should  damp out exponentially in time. This can be achieved by phenomenologically modifying Fick's law \req{fick}. Consider introducing a relaxation time scale $\tau_\pi$ so that \req{fick} is modified to
\begin{equation}
\vec{j} = -D \, \nabla \rho - \tau_\pi \, \p_t \vec{j}\ ,
\label{fick2}
\end{equation}	
One can think of the deformation to Fick's law described above as an example of the leading irrelevant operator that we include in the system. Now eliminating the current between \req{ctyeq} and \req{fick2} we find a hyperbolic equation, for
\begin{equation}
\frac{\p\rho}{\p t} - D \, \nabla^2 \rho -\tau_\pi\; \nabla \cdot\p_t {\vec j}=
 0 
\;\;\leadsto \;\;\frac{\p\rho}{\p t} - D \, \nabla^2 \rho +\tau_\pi\; \p_t^2 \rho= 0 \ .
\label{diffeq2}
\end{equation}	
Note  that we have used the fact that we are interested in  the lowest order terms in the gradient expansion and are therefore free to use the lower order equations of motion \req{diffeq} to simplify the dynamical equations. \req{diffeq2} has a finite velocity for the propagation of the density fluctuations, $v_\text{prop} = \sqrt{D/\tau_\pi}$. 

The toy model illustrates the basic idea involved in the  Israel-Stewart construction of ``causal relativistic hydrodynamics'' -- in their construction one adds a class of second derivative terms to the viscous relativistic hydrodynamics \req{viscous1}. It transpires that their construction actually ignores some of the two derivative operators allowed by the symmetries, see the discussion in  \citep{Baier:2007ix} in the context of  conformal fluids. 

However, as clearly described in \citep{Geroch:2001xs} this whole discussion is rather misleading. The essential point is that the effects of the higher order terms which are added to render the theory causal are of decreasing relevance deep into the hydrodynamic regime. From the effective field theory viewpoint emphasized here this is completely obvious -- the irrelevant operators have diminishing role to play as we move into the deep infra-red. Modes which manifest acausal behaviour are outside the long-wavelength regime and are not part of the hydrodynamic description. Curing this causality problem completely requires us to actually go back to the microscopic description of the theory. It is nevertheless interesting to ask what the systematic procedure to extend the relativistic viscous hydrodynamics is and we will address this question to second order in derivatives in the rest of these lectures. 

\section{Conformal fluids}
\label{s:confflu}

So far our discussion of fluid dynamics has been quite general and is valid for the hydrodynamic limit of any interacting Lorentz invariant quantum field theory. We now turn to a discussion of field theories which are conformally invariant.  After briefly reviewing the conformal transformation properties of the stress tensor, we turn to the construction of operators in the gradient expansion that transform homogeneously under conformal transformations. This will allow us to classify the set of operators that can appear in conformal hydrodynamics beyond leading order; for pure energy momentum transport this was first undertaken in \citep{Baier:2007ix}. We will then undertake a brief overview of an  extremely useful framework for discussing conformal fluids, the so called Weyl covariant formalism \citep{Loganayagam:2008is}. This will be useful when we construct gravitational duals to conformal fluids in \sec{s:gravity}.

\subsection{Weyl transformation of the stress tensor}
\label{s:weyl}

Consider a relativistic fluid on a background manifold $\CB_d$ with metric $g_{\mu \nu}$. 
We wish to consider a local Weyl rescaling of the metric and understand the transformation of the stress tensor. Given a Weyl transformation of the boundary metric
\begin{equation} 
 \begin{split}
 g_{\mu\nu} &= e^{2\phi}\, \widetilde{g}_{\mu\nu} \;\; \Rightarrow \;\; 
  g^{\mu\nu} = e^{-2\phi}\widetilde{g}^{\mu\nu} ,
\end{split}
\label{conftransf}
\end{equation}
it is clear that the velocity field appearing in the stress tenor transforms as 
\begin{equation}
 u^\mu = e^{-\phi}\, \widetilde{u}^\mu \ ,
 \label{velweyl}
\end{equation}	
which is a direct consequence of the normalization of the velocity field, $u^\mu \,u_\mu = -1$. It follows from \eqref{conftransf} and \req{velweyl} that the spatial projector transforms homogeneously as well: $P^{\mu\nu}=g^{\mu\nu}+u^{\mu}u^{\nu}=e^{-2\phi}\, \widetilde{P}^{\mu\nu}$.

We are interested in the behaviour of the stress tensor under conformal transformations. 
In general a tensor $\CQ$ with components $\CQ_{\mu_1 \cdots \mu_n}^{\nu_1 \cdots \nu_m}$ is said to be conformally invariant if it transforms homogeneously under Weyl rescalings of the metric, i.e., $\CQ = e^{-w\,\phi} \, \widetilde{\CQ}$ under \req{conftransf}.
The real number $w$ is the conformal weight of the tensor.  It is important to remember that the weight of a tensor operator under Weyl transformations depends on the index positions.\footnote{One could equivalently talk about the {\em invariant conformal dimension} which involves a shift by a simple linear combination of the number of upper and lower indices: 
$w_{\text{inv}} = w - n_{\text{lower}} + n_{\text{upper}}$. This follows from accounting for the weight of the metric.} We also should require that the dynamical equations satisfied by $\CQ$ remain invariant under conformal transformations.

It is easy to work out the constraints on the stress tensor using the dynamical equation at hand \req{cons1}. We find that first of all the conformal invariance requires that the stress tensor be traceless $T^\mu_\mu = 0$ and that it transforms homogeneously under 
Weyl rescalings of the metric with weight $d+2$
\begin{equation}
T^{\mu \nu} = e^{-(d+2) \,\phi}\, {\widetilde T}^{\mu \nu} \ .
\label{stenscale}
\end{equation}	
These statements follow from the conservation equation \req{cons1} for the stress tensor, written in the Weyl transformed frame. We refer the reader to Appendix D of \citep{Wald:1984gr} for a derivation of these results. The tracelessness condition in fact supplies the equation of state for conformal fluids. Going back to the ideal fluid \req{ideal1}, we find that 
\begin{equation}
T_\mu^\mu =0 \;\; \thus \;\; P = \frac{1}{d-1} \, \rho \ .
\label{confideal}
\end{equation}	
This relation between pressure and energy density fixes the speed of sound in conformal fluids to be a simple function of the spacetime dimension: $c_s = \frac{1}{\sqrt{d-1}}$. Likewise it is not hard to show the charge current transforms homogeneously with weight $d$ under conformal transformations 
\begin{equation}
J^\mu= e^{-d \,\phi}\, {\widetilde J}^\mu  . 
\label{ccurscale}
\end{equation}	

Another piece of data we need is the information of the scaling dimensions of the thermodynamic variables. It is for instance easy to see that the temperature  scales under the conformal transformation with weight $1$. Similarly it follows from the thermodynamic Gibbs-Duhem relation, $P+\rho = s\, T + \mu_I\, q_I$, that the chemical potentials $\mu_I$ also have weight $1$. Hence using \req{stenscale} and \req{conftransf}  it follows that the energy density is given by a simple scaling law (a natural extension of the Stefan-Boltzmann scaling to $d$ spacetime dimensions)
\begin{equation}
T  = e^{-\phi} \, \tilde{T} \;\; \thus \;\; \rho \sim T^{d}
\label{tempscale}
\end{equation}	
These results together imply that an ideal conformal fluid has a stress tensor 
\begin{equation}
\left(T^{\mu\nu}\right)_{\text{ideal}} = \alpha \, T^d\, \left(g^{\mu \nu} + d \, u^\mu \ u^\nu\right) \ , 
\label{idealc1}
\end{equation}	
where $\alpha$ is a dimensionless normalization constant which depends on the underlying microscopic CFT. 

To construct the fluid stress tensor at higher orders we simply need to enumerate the set of operators in our gradient expansion which transform homogeneously under \req{conftransf}. Let us consider the situation at first order in derivatives explicitly. Using the fact that the Christoffel symbols transform as \citep{Wald:1984gr}
\begin{equation*}
 \Gamma_{\lambda\mu}^{\nu}= \widetilde{\Gamma}_{\lambda\mu}^{\nu} + \delta^{\nu}_{\lambda}\, \partial_{\mu}\phi+ \delta^{\nu}_{\mu}\, \partial_{\lambda}\phi- \widetilde{g}_{\lambda\mu}\,\widetilde{g}^{\nu\sigma}\, \partial_{\sigma}\phi\ ,
\end{equation*}
we can show that the covariant derivative of $u^{\mu}$ transforms inhomogeneously:
\begin{equation}
 \nabla_{\mu}u^{\nu}=\partial_{\mu}u^{\nu}+\Gamma_{\mu\lambda}^{\nu}\, u^{\lambda}
  =e^{-\phi}\, \left[\widetilde{\nabla}_{\mu}\, \widetilde{u}^{\nu}+\delta^{\nu}_{\mu}\, \widetilde{u}^{\sigma}\, \partial_{\sigma}\phi- \widetilde{g}_{\mu\lambda}\, \widetilde{u}^{\lambda}\, \widetilde{g}^{\nu\sigma}\, \partial_{\sigma}\phi\right].
\end{equation}
This equation can be used to derive the transformation of various quantities of interest in fluid dynamics, such as the acceleration $a^\mu$, shear $\sigma^{\mu \nu}$, etc.. 
\begin{equation}
\begin{split}
\theta &= \nabla_{\mu}u^{\mu}
  =e^{-\phi}\, \left({\widetilde\nabla}_{\mu} {\widetilde u}^{\mu}+(d-1)\, \widetilde{u}^{\sigma}\, \partial_{\sigma}\phi\right) = e^{-\phi} \, \left(\widetilde{\theta} + (d-1)\, \widetilde{\DD} \phi \right) \ ,\\
 a^{\nu}&= \DD u^\nu = u^{\mu}\nabla_{\mu}u^{\nu}
  =e^{-2\phi}\left(\widetilde{a}^{\nu}+\widetilde{P}^{\nu\sigma}\, \partial_{\sigma}\phi\right),\\
 \sigma^{\mu\nu} &= P^{\lambda(\mu} \nabla_\lambda u^{\nu)}   - \frac{1}{d-1} \, P^{\mu\nu}\,\nabla_{\lambda}u^{\lambda} 
= e^{-3\,\phi} \; \widetilde{\sigma}^{\mu\nu},\\
\l^{\mu} &=  u_{\alpha}\,\epsilon^{\alpha \beta \gamma \mu}
\nabla_\beta u_{\gamma} = e^{-2 \,\phi}\, {\widetilde \l^\mu} \ ,
 \end{split}
 \label{fderdefs}
\end{equation}
where in the last equation we have accounted for the fact that all epsilon symbols
should be treated as tensor densities in curved space. The objects with correct tensor transformation properties scale as metric determinants \ie,    $\epsilon_{\alpha \beta \gamma \delta}  \propto \sqrt{-g}$, and  $\epsilon^{\alpha \beta \gamma \delta}\propto \frac{1}{\sqrt{-g}}$, from which it is easy to infer their scaling behaviour under conformal transformations; in particular, $\epsilon_{\alpha \beta\gamma \delta} = e^{4\phi} \; {\widetilde \epsilon}_{\alpha \beta \gamma \delta}$ and  $\epsilon^{\alpha \beta\gamma \delta} = e^{-4\phi} \;  {\widetilde \epsilon}^{\,\alpha \beta \gamma \delta}$. 

Armed with this information we can examine the terms appearing in the first order conformal fluid. From \req{fderdefs} we learn that the expansion $\theta$, transforms inhomogeneously implying that the coefficient of the bulk viscosity should vanish for a conformal fluid $\zeta = 0$.\footnote{Equivalently, this follows from the tracelessness of the stress tensor.} Likewise, for the charge current we can show that the contribution from the chemical potential and temperature should be in the combination $\mu_I/T$; the term involving gradient of the temperature $P^{\mu\nu}\,\nabla_\nu T$ transforms inhomogeneously under Weyl transformations requiring $\gamma_I = 0$. Therefore a conformal viscous fluid has a stress tensor which to first order in the gradient expansion
takes the form
\begin{eqnarray}
T^{\mu \nu}&=& \alpha\, T^d \, \left(g^{\mu \nu} +d\,  u^\mu \, u^\nu\right)  -2\, \eta \, \sigma^{\mu \nu} \ ,  \nonumber \\
J^\mu_I &=& q_I \, u^\mu -\varkappa_{IJ} \, P^{\mu\nu} \, \nabla_{\nu} \left(\frac{\mu_J}{T}\right) -  \mho_I\, \ell^\mu \ ,
\label{viscous2}
\end{eqnarray}	
where we have used the generalized Stefan-Boltazmann expression for the pressure.
We also can deduce the conformal properties of the transport coefficients; in particular,
\begin{equation}
\eta = T^{d-1}\, \hat{\eta}\left(\mu_I/T\right) .
\label{}
\end{equation}	
%

\subsection{Weyl covariant formulation of conformal fluid dynamics}
\label{s:weylcov}

It is straightforward to carry out this exercise to higher orders in the gradient expansion by again analyzing the transformation properties of the various operators under Weyl rescalings. This was carried out for uncharged fluids in \citep{Baier:2007ix} to second order and a detailed check of this for the specific second order stress tensor derived holographically is given in \citep{Bhattacharyya:2007lq}. However, it is useful to work a little more abstractly and exploit a little more of the structure of the conformal symmetry to present the answer compactly in a Weyl covariant form. We therefore will pause to review some details of this presentation following the beautiful work of \citep{Loganayagam:2008is}.

The basic point is as follows: when we are discussing a conformal field theory  and the associated hydrodynamic description on a background manifold $\CB_d$ we are not interested in the metric data on $\CB_d$, but rather on the conformal structure on this background manifold. We can exploit this fact to simplify the structure of the derivative expansion by directly working with the conformal class of metrics on $\CB_d$. For brevity, let us denote the background with this conformal class of metrics as $\left(\CB_d, \CC\right)$. 

On $\left(\CB_d, \CC\right)$ we will define a new derivative operator that keeps track of the Weyl transformation properties better. A key fact that we will exploit is that fluid dynamics comes equipped with a distinguished vector field, the velocity, which as emphasized earlier is just the timelike unit normalized eigenvector of the stress tensor.  We start by defining a torsionless connection called the Weyl connection, whose associated covariant derivative captures the fact that the metric transforms homogeneously under conformal transformations (with weight $-2$). In particular, the Weyl connection $\nabla^{\text{Weyl}}$ requires that for every metric in the conformal class $\CC$ there exists a connection one-form $\CA_\mu$ such that 
\begin{equation}
\nabla^{\text{Weyl}}_\alpha g_{\mu \nu}  = 2 \, \CA_\alpha \,g_{\mu\nu} \ .
\label{wcmet}
\end{equation}	
Given this derivative structure, we can go ahead and define a Weyl covariant derivative  $\WD_\mu$ as whose action on tensors transforming homogeneously with weight $w$ under i.e., $\CQ^{\mu\cdots}_{\nu \cdots} = e^{-w\, \phi} \,  \widetilde{\CQ}^{\mu\cdots}_{\nu \cdots}$  is given by
\begin{eqnarray}
\WD_\lambda \CQ^{\mu\cdots}_{\nu \cdots} &\equiv& \nabla_\lambda \CQ^{\mu\cdots}_{\nu \cdots} + w\, \CA_\lambda \,\CQ^{\mu\cdots}_{\nu \cdots}\nonumber \\
&&  +\left( g_{\lambda\alpha}\, \CA^\mu - \delta^\mu_\lambda \, \CA_\alpha - \delta^\mu_\alpha \, \CA_\lambda\right) \,  \CQ^{\alpha\cdots}_{\nu \cdots} + \cdots \nonumber \\
&&- \left( g_{\lambda\nu}\, \CA^\alpha - \delta^\alpha_\lambda \, \CA_\nu - \delta^\alpha_\nu \, \CA_\lambda\right) \, \CQ^{\mu\cdots}_{\alpha \cdots} -\cdots \ .
\label{weylcd}
\end{eqnarray}	
The nice thing about this definition is that $\WD_\lambda  \CQ^{\mu\cdots}_{\nu \cdots} = 
 e^{-w\, \phi} \, \widetilde{\WD}_\lambda \widetilde{\CQ}^{\mu\cdots}_{\nu \cdots}$,  i.e., the Weyl covariant derivative of a conformally invariant tensor transforms homogeneously with the same weight as the tensor itself.

One can view the Weyl covariant derivative as being determined in terms of the Weyl connection via $\WD_\mu = \nabla^{\text{Weyl}}_\mu + w \, \CA_\mu$. From this description it follows immediately that the Weyl connection is  metric compatible $\WD_\alpha g_{\mu\nu} = 0$ which follows from \req{wcmet} and the fact that $w=-2$ for $g_{\mu\nu}$. In addition, in fluid dynamics we will require that the Weyl covariant derivative of the fluid velocity be transverse and traceless, i.e., 
\begin{equation}
u^\alpha \, \WD_\alpha u^\mu = 0 \ , \qquad \WD_\alpha u^\alpha = 0 \ .
\label{}
\end{equation}	
These conditions enable to uniquely determine the connection one-form $\CA_\mu$ to be the the distinguished vector field 
\begin{equation}
\CA_\mu = u^\lambda\, \nabla_\lambda u_\mu - \frac{1}{d-1}  \, u_\mu 
\, \nabla^\lambda u_\lambda \equiv a_\mu - \frac{1}{d-1}\, \theta \, u_\mu  \  , 
\label{cadef}
\end{equation}	
built from the velocity field of fluid dynamics.

One can rewrite the various quantities appearing in the gradient expansion of the stress 
tensor  in this Weyl covariant notation. For instance at first order in derivatives we have the objects constructed from the velocity field
\begin{equation}
\sigma^{\mu\nu} = \WD^{(\mu} u^{\nu)} \ , \qquad 
\omega^{\mu\nu} = -\WD^{[\mu} u^{\nu]} \ ,
\label{udecdefs2}
\end{equation}
both of which can be seen to have weight $w =3$.  The dynamical equations of fluid dynamics \req{cons1} can be recast into the Weyl covariant form. For instance,  stress tensor conservation can be easily seen to be $\WD_\mu \, T^{\mu \nu} = 0$; for 
\begin{eqnarray}
\WD_\mu T^{\mu \nu} &=& \nabla_{\mu}\ T^{\mu \nu} +w \, \CA_\mu \, T^{\mu \nu} + 
\left( g_{\mu\alpha}\, \CA^\mu - \delta^\mu_\mu\, \CA_\alpha - \delta^\mu_\alpha \, \CA_\mu \right) \, T^{\alpha \nu} +  \left(g_{\mu\alpha}\, \CA^\nu - \delta^\nu_\mu \, \CA_\alpha - \delta^\nu_\alpha \, \CA_\mu\right) \, T^{\mu \alpha} \nonumber \\
&=& \nabla_\mu \, T^{\mu \nu}+ (w -d -2)\, \CA_\mu \, T^{\mu \nu} - \CA^\nu\, T^\mu_\mu
\nonumber \\
&=& \nabla_\mu T^{\mu \nu}  \  ,
\label{}
\end{eqnarray}	
where we used the conformal weight $w=d+2$ of the stress tensor and the tracelessness condition to remove the inhomogeneous terms. For CFTs on curved manifolds one has the possibility of encountering a conformal anomaly in even spacetime dimensions;  incorporating the trace anomaly $W$ we can write the fluid dynamical equation as 
\begin{equation}
\WD_\mu T^{\mu\nu} = 
\nabla_\mu T^{\mu\nu}  + \CA^\nu \, \left(T^\mu_\mu - W \right) =0 \ .
\label{}
\end{equation}	
%

\subsection{Non-linear conformal fluids}
\label{s:wsec}

We will now discuss the general conformal fluids to second order in derivative expansion using the Weyl covariant formalism, with an aim to cast fluid mechanics in manifestly conformal language. In order to achieve this we need to write down the set of two derivative operators that transform homogeneously under conformal transformations.  These operators involve either two derivatives acting on the dynamical degrees of freedom or terms involving squares of first derivatives.  We now proceed to enumerate these operators: 

For operators built out solely from the fluid velocity, one has the following terms which explicitly have two derivatives 
\begin{equation}
\begin{split}
\WD_{\mu}\WD_{\nu} u^\lambda &= \WD_{\mu} {\sigma}_{\nu}{}^{\lambda} + \WD_{\mu}{\omega}_{\nu}{}^{\lambda} = e^{-\phi}\widetilde{\WD}_{\mu}\widetilde{\WD}_{\nu}\tilde{u}^{\lambda}\\
\WD_{\lambda} {\sigma}_{\mu\nu} &=  e^{\phi}\, \widetilde{\WD}_{\lambda} \widetilde{\sigma}_{\mu\nu} \ , \qquad 
\WD_{\lambda} {\omega}_{\mu\nu} =e^{\phi} \,\widetilde{\WD}_{\lambda} \widetilde{\omega}_{\mu\nu}  \ . 
\end{split}
\label{vel2der}
\end{equation}
In addition we can have terms which have the combinations:
\begin{equation}
\sigma^\mu_{\;\alpha} \, \sigma^{\nu\alpha}  = e^{-4\,\phi}\, \widetilde{\sigma} ^\mu_{\;\alpha} \, \widetilde{\sigma} ^{\nu\alpha} \ ,\qquad 
\omega^\mu_{\;\alpha} \, \omega^{\nu\alpha} =  e^{-4\,\phi}\,\widetilde{\omega}^\mu_{\;\alpha} \, \widetilde{\omega}^{\nu\alpha} \ , \qquad 
\sigma ^\mu_{\;\alpha}\,\omega^{\nu\alpha} = e^{-4\,\phi}\,\widetilde{\sigma} ^\mu_{\;\alpha}\,\widetilde{\omega}^{\nu\alpha} \ , 
\label{vel1dersq}
\end{equation}	
which involve the squares of the first derivative operators. 

In addition to the operators built out of the velocity field we should also consider terms which involve the temperature $T$ and various chemical potentials $\mu_I$. Recalling that both of these have weight $1$ under conformal transformations, i.e.,  $T=e^{-\phi}\, \widetilde{T}$ and $\mu_I=e^{-\phi}\,\tilde{\mu}_I$, we can write down the relevant operators involving no more than two derivatives:  
\begin{equation}
\begin{split}
\WD_\mu\left(\frac{\mu_I}{T}\right) &= \widetilde{\WD}_\mu  \left(\frac{\widetilde{\mu}_I}{\widetilde{T}}\right) \ ,\qquad \WD_\mu T = e^{-\phi} \widetilde{\WD}_\mu \widetilde{T} \\
\WD_\lambda\WD_\sigma\left(\frac{\mu_I}{T}\right) &= \widetilde{\WD}_\lambda\widetilde{\WD}_\sigma\left(\frac{\widetilde{\mu}_I}{\widetilde{T}}\right) \, \qquad 
\WD_\lambda\WD_\sigma T = e^{-\phi}\, \widetilde{\WD}_\lambda\widetilde{\WD}_\sigma\widetilde{T}  \ .
\end{split}
\label{Tmuders}
\end{equation}

To complete the classification of the various tensors that can be constructed at the second derivative level, we need to study the curvature tensors that appear via the commutators of two covariant derivatives. These appear because the fluid couples to the background metric $g_{\mu\nu}$ on the manifold $\CB_d$ and the curvature terms arise at second order in derivatives. These terms will be of particular importance in the holographic context; for example in deriving the fluid dynamical behaviour of $\CN =4$ SYM on $\Sp^3 \times \R$ we encounter curvature terms. We review the  definition of the curvature tensors in the Weyl  covariant formalism in \App{a:wcurve}. For the present we just record the symmetric traceless tensor involving two derivatives of the background metric on $\CB_d$, 
\begin{equation}
C_{\mu\alpha\nu\beta}\,u^\alpha \,u^\beta  = \widetilde{C}_{\mu\alpha\nu\beta}\,\tilde{u}^\alpha \,\tilde{u}^\beta 
\end{equation}	
where $C_{\alpha \beta \mu\nu}$ is the Weyl tensor see \req{wweyl}.

\subsection{The non-linear conformal stress tensor}
\label{s:nln4}

We are now in a position to discuss the non-linear hydrodynamics for conformal fluids. 
For the sake of brevity we will confine attention to fluids which have no conserved charges. This is equivalent to working in the canonical ensemble and focussing just on energy momentum transport. The results we derive here will be universal for a wide class of conformal fluids.

Let us classify all the operators which are Weyl invariant at various orders in the derivative expansion. The results of the previous section \sec{s:wsec} can be summarized as follows: at first  two orders in derivatives the set of  symmetric traceless tensors which transform homogeneously under Weyl rescalings are given to be\footnote{Various papers in the literature seem to use slightly different conventions for the normalization of the operators. We will for convenience present the results in the normalizations used initially in  \citep{Baier:2007ix}. This is the  source of the  factors of $2$ appearing in the definition of the tensors $\bCT_i$, see \citep{Bhattacharyya:2007lq} for a discussion.}
\begin{equation}
\begin{aligned}
\text{First order}: \;\;&  \sigma^{\mu\nu}  &\nonumber \\
\text{Second order}: \;\; &\bCT_1^{\mu\nu} =2\, u^\alpha \, \WD_\alpha \sigma_{\mu\nu} \ , \quad 
 \bCT_2^{\mu\nu} =C_{\mu\alpha\nu\beta}\,u^\alpha \,u^\beta \ , \nonumber \\
 \qquad&\bCT_3^{\mu\nu}  =4\,\sigma^{\alpha\langle\mu}\, \sigma^{\nu\rangle}_{\ \alpha}  \ , \quad 
  \bCT_4^{\mu\nu} =  2\, \sigma ^{\alpha\langle\mu}\, \omega^{\nu\rangle}_{\ \alpha}  \ , \quad  \bCT_5^{\mu\nu}=   \omega ^{\alpha\langle\mu}\, \omega ^{\nu\rangle}_{\ \alpha} 
\end{aligned}
\label{winv2der}
\end{equation}	
where we have introduced a notation for the second derivative operators which will be useful to write compact expressions for the stress tensor below. Armed with this data  we can immediately  write down the general contribution to the stress tensor as:
\begin{eqnarray}
\Pi_{(1)}^{\mu\nu} &=& -2\, \eta\,\sigma^{\mu\nu} \nonumber \\
\Pi_{(2)}^{\mu\nu} &=&  \tau_\pi\,\eta\, \bCT_1^{\mu\nu} + \kappa\,\bCT_2^{\mu\nu} + \lambda_1\, \bCT_3^{\mu\nu} + \lambda_2\, \bCT_4^{\mu\nu} + \lambda_3\, \bCT_5^{\mu\nu} \  .
\label{fld2}
\end{eqnarray}	
There are therefore have six transport coefficients $\eta$, $\tau_\pi$, $\kappa$, $\lambda_i$ for $i = \{1,2, 3\}$, which characterize the flow of a non-linear viscous fluid.

For a fluid with holographic dual using the explicit result of the gravitation solution \req{formmet} (see \sec{s:gravnl}) and the holographic stress tensor \citep{Henningson:1998gx,Balasubramanian:1999re} we find explicit values for the transport coefficients. In particular, for the $\CN=4$ SYM fluid one has\footnote{The result for generic $\CN =1$ superconformal field theories which are dual to gravity on \AdS{5} $\times X_5$ are given by simply replacing $\frac{N^2}{8 \pi^2}$ by the corresponding central charge of the SCFT.} \citep{Baier:2007ix,Bhattacharyya:2007lq}
\begin{equation}
\begin{aligned}
&\eta = \frac{N^2}{8 \pi} \, \left(\pi  T\right)^3 \;\;  \thus \;\;\frac{\eta}{s} = \frac{1}{4\pi}  \ , \\
&\tau_\pi = \frac{2 - \ln 2}{2\pi\, T} \ , \qquad \kappa = \frac{\eta}{\pi \,T} \\
& \lambda_1 = \frac{\eta}{2\pi\, T} \ , \qquad
\lambda_2 = \frac{\eta\, \ln 2}{\pi\, T}\ , \qquad \lambda_3=0 \ . 
\end{aligned} 
\label{transp4d}
\end{equation}	
where in the first line we have used the standard entropy density for thermal $\CN =4$ SYM at strong coupling $s = \frac{\pi^3}{2}\,N^2 \, T^3$ to exhibit the famous ratio of shear viscosity to entropy density \citep{Kovtun:2004de}.  

The general analysis in \AdS{d+1} of carried out in \citep{Haack:2008cp, Bhattacharyya:2008mz} in fact allows us to write down the transport coefficients described above in arbitrary dimensions in nice closed form.\footnote{For $d=3$ the general results were initially derived in \citep{VanRaamsdonk:2008fp}. See also \citep{Natsuume:2008iy} for determination of some of these coefficients in $d =3$ and $d =6$.} The transport coefficients for conformal fluids in $d$-dimensional boundary $\CB_d$ are:
\begin{equation}
\begin{aligned}
&\eta = \frac{1}{16 \pi\, G_N^{(d+1)}} \, \left(\frac{4\pi}{d}\,   T\right)^{d-1} \;\;  \thus \;\;\frac{\eta}{s} = \frac{1}{4\pi}  \ , \\
&\tau_\pi = \frac{d}{4\pi\, T} \, \left[1 + \frac{1}{d}\, \text{Harmonic}\left(\frac{2}{d} -1\right) \right] , \qquad \kappa = \frac{d}{2\pi\, (d- 2)} \,\frac{\eta}{T} \\
& \lambda_1 = \frac{d}{8\pi}\, \frac{\eta}{T} \ , \qquad
\lambda_2 = \frac{1}{2\pi}\,  \text{Harmonic}\left(\frac{2}{d} -1\right) \, \frac{\eta}{T}\ , \qquad \lambda_3=0 \ . 
\end{aligned} 
\label{transpgend}
\end{equation}	
where $\text{Harmonic}(x)$ is the harmonic number function.\footnote{The harmonic number function may be in fact be re-expressed in terms of the digamma function, or more simply as 
$$\text{Harmonic}(x) = \gamma_e + \frac{\Gamma'(x)}{\Gamma(x)} \ , $$ 
where $\gamma_e$ is Euler's constant;
see \citep{Sondow:2009ng} for a discussion of its properties.}

\section{Non-linear fluid dynamics from gravity}
\label{s:gravity}

Our discussion of fluid dynamics has thus far been very general and in fact is valid for any interacting field theory. As we have seen one can distill the information in this low energy effective description into the specification of a finite number of transport coefficients. The transport coefficients can in principle be determined from the microscopic field theory. However, if we are interested in systems which are intrinsically strongly coupled then we need to find a technique to extract the non-perturbative values of the transport coefficients.  In the rest of these lectures we will focus on such strongly coupled scenarios. As is well known, for a class of such theories one can exploit the AdS/CFT correspondence to extract the hydrodynamical parameters and this will be our prime focus.

Let us consider a $d$ dimensional field theory on a background manifold $\CB_d$ which is holographically dual to string theory on an asympotically \AdS{d+1} spacetime (which is perhaps warped with some internal compact manifold to be a critical string background). The prototype example which we should keep in mind is the classic duality between Type IIB string theory on \AdS{5} $\times \Sp^5$ and the dynamics of $\CN =4$ SYM. Following standard parlance we will refer to the field theory as living on the boundary of the asymptotically \AdS{} spacetime.\footnote{For four dimensional boundary field theories  we can replace the $\Sp^5$ by a Sasaki-Einstein manifold $X_5$ such as  $T^{1,1}$ or $L^{p,q,r}$. In these cases one recovers $\CN =1$ quiver superconformal gauge theories on the boundary.} In the example of $\CN =4 $ SYM we have two dimensionless parameters in the field theory, the 't Hooft coupling $\lambda$ and the rank of the gauge  group $N$ (we will consider unitary gauge group $SU(N)$). In the limit of large $N$ and large $\lambda$ the field theory dynamics is captured by classical gravity as the dual string coupling is small and one moreover has a macroscopically large \AdS{5} spacetime. 

In the planar large $N$ limit, the field theory dynamics truncates to the dynamics of single trace operators $\CO =  \frac{1}{N} \, {\rm Tr} \left(\Psi\right)$ where $\Psi$ is a collection of the basic fields appearing in the Lagrangian. For $\CN =4$ SYM  one has the bosonic R-charge scalars transforming in the vector representation of the $SU(4)$ R-symmetry, the gauge fields,  and fermions; schematically $\Psi = \{X^i, D_\mu, \psi_\alpha\}$. The  dynamics in the single trace sector in the planar limit is classical due to large $N$ factorization, but the general structure is hard to derive even for the particularly symmetric case of $\CN =4$ SYM. From a field theory viewpoint issues such a locality of this effective classical description are rather mysterious. However, once we pass to the holographic dual we encounter a manifestly local description in terms of two derivative gravity coupled to a bunch of matter fields in an asymptotically \AdS{5} spacetime. In making this statement we are assuming that we have performed a Kaluza-Klein reduction of the Type IIB supergravity fields over the compact $\Sp^5$ leading to an infinite tower of massive fields coupled to the gravitational degrees of freedom. 

The general structure of this effective five-dimensional lagrangian is not only complicated but it also depends on the details of the internal space. Were one to replace the $\Sp^5$ by a Sasaki-Einstein five manifold $X_5$ one would end up with a different effective description corresponding to a different field theory fixed point in four dimensions. However, there is a {\em universal sub-sector} of Type IIB supergravity which we can focus on -- this is just the sector of solely gravitational dynamics in \AdS{5} i.e., we set all the Kaluza-Klein harmonics of the graviton modes on $\Sp^5$ and other matter degrees of freedom consistently to zero. We will restrict attention to this sub-sector which corresponds in the dual field theory to focussing on just the dynamics of the energy-momentum tensor.

\subsection{The universal sector: gravity in \AdS{d+1}}
\label{s:univgravity}

As discussed above we will concentrate on pure gravitational dynamics in asymptotically \AdS{} spacetimes. This in particular allows us to work without loss of generality in arbitrary dimensions as the form the gravitational action is independent of the number of spacetime dimensions. Let us therefore consider starting with a string or M-theory background of the form \AdS{d+1} $\times X$ where $X$ is some compact internal manifold ensuring that one has a consistent string/M-theory vacuum.\footnote{We will be interested in $d> 2$. As discussed in \citep{Bhattacharyya:2008xc,Bhattacharyya:2008mz} there is no interesting hydrodynamic limit for a $1+1$ dimensional CFT. A conserved traceless stress tensor is simply made up of left and right moving waves which propagate with no dissipation.}  The universal sector of this theory which we focus on is the dynamics of Einstein gravity with a negative cosmological constant, i.e., 
\begin{equation}
\CS_{\text{bulk}} = \frac{1}{16\pi\,G_N^{(d+1)}}\,\int\, d^{d+1}x\, \sqrt{-G} \,\left(R - 2\,\Lambda\right) \ .
\label{ehact}
\end{equation}	
 With a particular choice of units  ($R_{AdS} = 1$)  Einstein's equations are
given by\footnote{We use upper case Latin indices $\{M,N, \cdots\}$ to denote bulk directions, while lower case Greek indices $\{\mu ,\nu, \cdots\}$ refer to field theory or boundary directions. Finally, we use lower case Latin indices $\{i,j,\cdots\}$ to denote the spatial directions in the boundary.}
\begin{equation}\label{ein} \begin{split}
&E_{M N}= R_{M N} - \frac{1}{2} \, G_{M N} R- \frac{d(d-1)}{2}\, G_{MN}=0\\
\implies &R_{M N} + d\, G_{M N}=0, \qquad R=- d(d+1) . 
\end{split}
\end{equation}

Of course the equations \eqref{ein} admit \AdS{d+1} solutions, which correspond to the vacuum state of the dual field theory. Recall that global \AdS{d+1} has as its boundary the Einstein static universe, $\R \times \Sp^{d-1}$. We are free to consider other choices of boundary manifolds $\CB_d$; for instance to discuss field theory on Minkowski space $\R^{d-1,1}$ we would focus on the Poincar\'e patch of \AdS{d+1}. Given a metric $g$ on the boundary $\CB_d$ we have the bulk geometry to zeroth  order in the Fefferman-Graham expansion given by\footnote{In fact, if $g_{\mu\nu}$ is Ricci flat then \req{ads1} satisfies the Einstein's equations \req{ein}. Similar constructions can be made for Einstein metrics on the boundary, albeit with a different warping function i.e., $ds^2 = \frac{1}{z^2}\,\left(dz^2 + F(z)\, g_{\mu\nu}\, dx^\mu\,dx^\nu\right)$ for an appropriate choice of $F(z)$. For instance $F(z) =(1+\frac{z^2}{4})^2$ for an Einstein manifold with negative curvature.}
\begin{equation}
ds^2 = \frac{1}{z^2}\,\left(dz^2 + g_{\mu\nu}\, dx^\mu\,dx^\nu\right) .
\label{ads1}
\end{equation}	
We will refer to these spacetimes as asymptotically \AdS{d+1} although this terminology is strictly speaking incorrect in a strict general relativistic sense.  We will return to the issue of how to find solutions to \req{ein} with this prescribed boundary metric, after discussing some issues related to fluid dynamics.

We are interested in the description of the field theory in the canonical ensemble and in particular on situations where the field theory only attains local thermal equilibrium. There is usually an interesting phase structure for the field theories on non-trivial boundary manifolds $\CB_d$. Usually this arises from the fact that one can have dimensionless ratios of length scales arising from the non-trivial geometry of the background. The classic example is $\CB_d = \R \times \Sp^{d-1}$ where the low temperature phase is described to be the confined phase with $\ord{1}$ free energy while the high temperature phase is the deconfined phase with $\ord{N^2}$ free energy \citep{Witten:1998zw}. The former phase is dual to a thermal gas in \AdS{d+1} while latter phase has a geometric description in terms of a Schwarzschild black hole in \AdS{d+1}. 

To discuss hydrodynamics we need to be in the long-wavelength regime which is achieved only in the deconfined phase at high temperatures \citep{Bhattacharyya:2007vs}.  A simple way to see this is to note that in conformal field theories the phase structure is determined by the dimensionless ratio of length scales -- if $\CB_d$ has curvature scale $R_c$ and we are interested in the canonical ensemble at temperature $T$, then the phase structure depends on $R_c \,T$. However, since the mean free path of the system $\ell_{\text{mfp}} \sim 1/T$ from conformal invariance, we see that in order  for the gradient expansion to be valid we require that $R_c\, T \gg 1$. This can equivalently be interpreted as the statement that variations in the curvature of the background are small in units of the local temperature which means that we can approximate the  boundary metric to be locally flat. This discussion is in perfect accord with the fact that for the CFT on Minkowski space the absence of any length scales in the background means that one has a trivial phase structure -- the field theory is always deconfined on $\R^{d-1,1}$. 

Therefore to construct the dual of hydrodynamics on a curved boundary manifold $\CB_d$ we can use as our starting point a flat boundary and systematically account for curvature terms as we proceed in the gradient expansion. In fact, the leading order viscous hydrodynamics is insensitive to the boundary curvature terms which only show up at second order in derivatives. 

\subsection{Preliminaries: Schwarzschild black holes in \AdS{d+1}}
\label{s:sads}
Let us consider the geometry dual to thermal field theory on Minkowski space, which is given by the planar \SAdS{d+1} black hole which in Schwarzschild type coordinates is given by
\begin{equation}
\begin{split}
ds^2 &= -r^2\, f(b\,r) \, dt^2 + \frac{dr^2}{r^2\, f(b\,r) }+ r^2\, \delta_{ij} \, dx^i\, dx^j \ ,\\
f(r) &= 1-\frac{1}{r^d} \ .
\end{split}
\label{sads1}
\end{equation}	
While this is a one-parameter family of solutions labeled by the horizon size $r_+$ which sets the temperature of the black hole
\begin{equation}
T= \frac{d}{4\pi\,b} \  ,
\label{Tschw}
\end{equation}	
it is easy to generate a $d$ parameter family of solutions by boosting the solution along the 
translationally invariant spatial directions $x^i$, leading to a solution:\footnote{The indices in the boundary  are  raised and lowered with the Minkowski metric \ie, $u_{\mu} = \eta_{\mu \nu} \, u^\nu$. }
\begin{equation}
ds^2 =  \frac{dr^2}{r^2\, f(b\,r) }+ r^2\, \left( -f(b\,r)\, u_\mu \,u_\nu +P_{\mu\nu}\right)\, dx^\mu \, dx^\nu 
\label{sads2}
\end{equation}	
with
\begin{equation} \label{defun} \begin{split}
u^v&=\frac{1}{\sqrt{1-\beta^2}} \\
u^i&=\frac{\beta_i}{\sqrt{1-\beta^2}} \ ,
\end{split}
\end{equation}
where the temperature  $T$ and velocities $\beta_i$ are all constants with $\beta^2 = \beta_j \, \beta^j$, and $ P^{\mu \nu}= u^\mu u^\nu +\eta^{\mu \nu} $
is the projector onto spatial directions.  These solutions are generated by a simple coordinate transformation from \req{sads1} and can be understood physically as follows. The isometry group of \AdS{d+1} is $SO(d,2)$. The Poincare algebra plus dilatations form a distinguished subalgebra of this group. The rotations $SO(d)$ and translations $\R^{1,d-1}$ that belong to this subalgebra annihilate the static black brane solution \req{sads1} in \AdS{d+1}. This is manifest from the symmetries of the background \req{sads1}.  However, the remaining symmetry generators, which are the  dilatations and boosts, act nontrivially on this brane, generating a $d$ parameter set of black hole solutions. The parameters which characterize the bulk solution are precisely the basic hydrodynamical degrees of freedom, viz., temperature and the velocity
of the black hole. 

The boosted black hole  \req{sads2} is an asymptotically \AdS{d+1} solution which has a holographic stress tensor on the boundary. It is in fact not hard to see that the stress tensor for this solution is the ideal conformal fluid stress tensor \req{idealc1} with a particular value  of the normalization constant $\alpha = \frac{\pi^d}{16\pi\,G_N^{(d+1)}}$. This is not surprising as the solution is stationary and therefore corresponds to the global thermal equilibrium. To describe hydrodynamics we should perturb the system away from global equilibrium. Based on our discussion in \sec{s:fluids} it is natural to consider a situation where the thermodynamic variables vary along the boundary directions. This can be simply achieved by promoting the parameters $b$, $\beta_i$ to functions of the boundary coordinates and furthermore letting the boundary metric vary  to account for curvature couplings. As long as the variations are of large wavelength we can work in the effective field theory framework and construct a solution order by order in boundary derivative expansion.

Roughly speaking, our construction may be regarded as  the `Chiral Lagrangian' 
for brane horizons. We have hitherto discussed how the dilatation and boost generators of the conformal algebra act on the space of black hole solutions which is described by a point in $d$ dimensional parameter space. Our construction effectively promotes these parameters to `Goldstone fields' (or perhaps more accurately collective coordinate fields) and determines the effective dynamics of these collective coordinate fields, order by order in the derivative expansion, but making no assumption about amplitudes.\footnote{Another useful way to think about the hydrodynamic description in gravity language is to view it as the collective field theory of the lowest quasi-normal mode, the mode with vanishing frequency at zero momentum, of the black hole geometry.}
\subsection{Regularity and choice of coordinate chart}
\label{s:regcoord}

Whilst it is straightforward to promote the parameters $b$ and $\beta_i$  appearing in \req{sads2} to functions depending on $(t,x^i)$, there is a subtlety we need to consider. The issue is that the Schwarzschild coordinates used to write the metric in \req{sads2} are not regular on the future horizon. We would like to work with coordinates which are manifestly regular everywhere except for the singularity at the origin $r =0$ in \req{sads2}, for we expect that the class of stress tensors that are fluid dynamical in nature should be special from a general relativistic viewpoint (we will shortly review why). In fact, we will show that  the fluid dynamical stress tensors give rise to regular black hole spacetimes as their holographic duals. 

One can motivate the special nature of fluid dynamical form for the stress tensor by the following observation: consider the Fefferman-Graham form of the \AdS{d+1} metric \req{ads1}. It is well known that in order to find an asympotically \AdS{d+1} spacetime with a prescribed boundary $\CB_d$, one has to give in addition to the metric $g_{\mu\nu}$ on $\CB_d$ another piece of data which corresponds to the boundary stress tensor. Armed with $g_{\mu\nu}$ and $T_{\mu\nu}$ one can construct a bulk solution order by order in a perturbative expansion in the Fefferman-Graham radial variable $z$. To leading order the solution is simply 
\begin{equation}
ds^2 = \frac{1}{z^2}\, \left(dz^2  + \left(g_{\mu\nu} + a \, z^d \, T_{\mu\nu}\right) \, dx^\mu \, dx^\nu\right) . 
\label{}
\end{equation}	
This scheme for constructing a bulk spacetime with prescribed boundary data is well developed in the formalism of holographic renormalization \citep{deHaro:2000xn}. However, this scheme is not likely to generically reproduce regular bulk spacetimes. To see why one just needs to a do a simple degrees of freedom counting. A  traceless, symmetric  stress tensor  on $\CB_d$ has $\frac{d(d+1)}{2}-1$ degrees of freedom. But the dynamical equations of motion are simply the conservation equations \req{cons1} which are just $d$ equations leading to a vastly underdetermined system when $d > 2$. However, a fluid dynamical stress tensor is a special class of conserved stress tensors for it is described by precisely $d$ degrees of freedom, the temperature and velocity.\footnote{Similar arguments can be given if we want to consider fluids which carry conserved global charges.} 

In order to make regularity manifest, we will describe how to construct gravitational black hole solutions dual to arbitrary fluid flows using a coordinate chart that is regular on the future horizon.\footnote{For a discussion on the Fefferman-Graham coordinates and regularity see \citep{Gupta:2008th}.} We work with a set of generalized Gaussian null coordinates which are constructed with the aim of having the putative horizon  located at some hypersurface $r(x^\mu) = r_H(x^\mu)$.\footnote{We will determine explicitly the location of the horizon after sketching the construction of the solution.} So as the starting point for our analysis we consider the boosted planar \SAdS{d+1} black hole solution: 
\begin{equation}
ds^2 =-2\, u_{\mu}\, dx^{\mu} dr -r^2\, f(b\,r)\, u_{\mu}\, u_{\nu}\, dx^{\mu}dx^{\nu} + r^2\, P_{\mu\nu}\, dx^{\mu} dx^{\nu} \ , 
\label{boostedbrane}
\end{equation}
where we have written the metric in ingoing Eddington-Finkelstein coordinates. We should note that it is possible to recast \req{boostedbrane} in a Weyl covariant form when the boundary metric on $\CB_d$ is curved -- we have \citep{Bhattacharyya:2008mz}:
\begin{equation}
ds^2 =-2\, u_{\mu}\, dx^{\mu} \,\left(dr+ r\,\CA_\mu\, dx^\mu\right) +r^2\,\left(1- f(b\,r)\right)\, u_{\mu}\, u_{\nu}\, dx^{\mu}dx^{\nu} + r^2\, g_{\mu\nu}\, dx^{\mu} dx^{\nu} \ .
\label{bbranew}
\end{equation}	

\begin{figure}[h!]
 \begin{center}
 \includegraphics[scale=0.75]{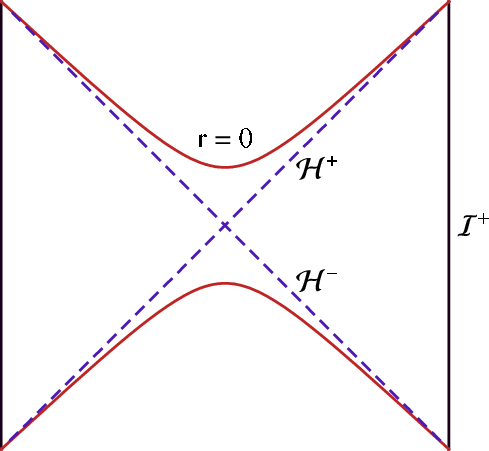}
\hspace{1.5cm}
 \includegraphics[scale=0.33]{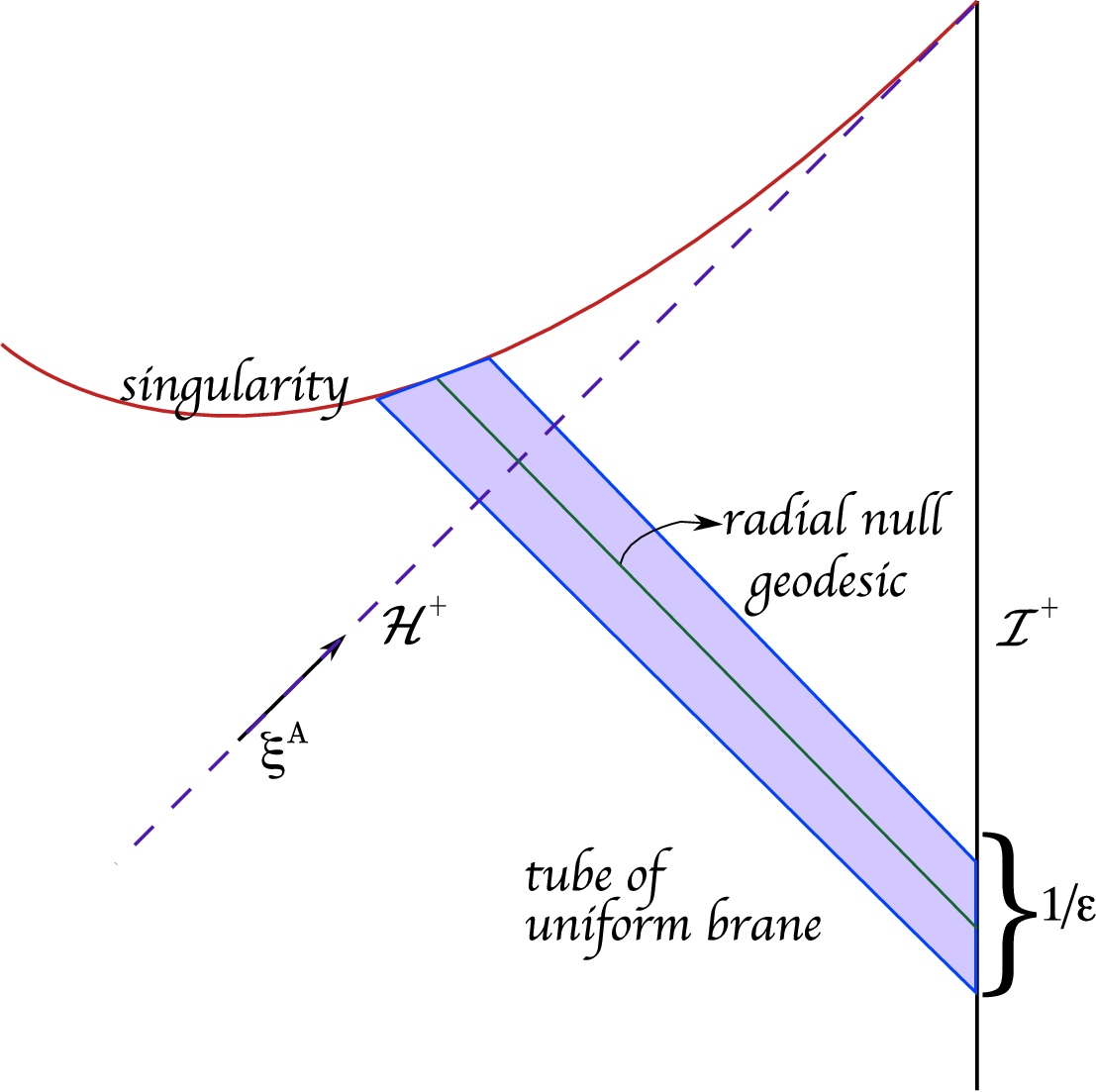} 
\end{center}
\caption{Penrose diagram of the uniform black brane and the causal structure of the spacetimes dual to fluid mechanics illustrating the tube structure. The dashed line in the second figure denotes the future event horizon, while the shaded tube indicates the region of spacetime over which the solution is well approximated by a tube of the uniform black brane.}
\label{PDtube}
\end{figure}

The main rationale behind switching to these Eddington-Finkelstein coordinates apart from making issues of regularity more transparent, is that they provide a clear physical picture of the locally equilibrated fluid dynamical domains in the bulk geometry. The boundary domains where local thermal equilibrium is attained in fact extend along ingoing radial null geodesics into the bulk. So a given boundary domain corresponds to an entire tube of width set by the scale of variation in the boundary, see \fig{PDtube} for an illustration. In the  Eddington-Finkelstein coordinates one just has to patch together these tubes to obtain a solution to Einstein's equations and moreover this patching can be done order by order in boundary derivatives, just as in fluid dynamics.  We now proceed to outline a perturbation scheme which allows us to construct the desired gravity solution dual to fluid dynamics. 

As the starting point consider the metric \eqref{boostedbrane} with the constant parameter $b$
and the velocities $\beta_i$ replaced by slowly varying functions
$b(x^\mu ), \beta_i(x^\mu)$ of the boundary coordinates.
\begin{equation}\label{boostedg0}
ds^2 =-2\,u_{\mu}(x^\alpha)\,dx^{\mu} \,dr -r^2\, f\left(b(x^{\alpha})\,r\right) \,
u_{\mu}(x^\alpha) \, u_{\nu} (x^\alpha) \, dx^{\mu}\, dx^{\nu} +
r^2\, P_{\mu\nu}(x^\alpha) \, dx^{\mu} \, dx^{\nu} \ .
\end{equation}
Generically, such a metric (we will denote it by  $G^{(0)}(b(x^\mu), \beta_i(x^\mu)$)
is not a solution to Einstein's equations. Nevertheless it has two attractive
features. Firstly, away from $r=0$, this deformed metric is everywhere
non-singular. This pleasant feature is tied to our use of
Eddington-Finkelstein coordinates. Secondly, if all derivatives of the parameters $b(x^\mu)$ and $\beta_i(x^\mu)$ are small, $G^{(0)}$ is ``tubewise'' well approximated by a boosted black brane. Consequently, for slowly varying functions $b(x^\mu)$,  $\beta_i(x^\mu)$, it might seem intuitively plausible that \eqref{boostedg0} is a good approximation to a true solution of Einstein's equations with a regular event horizon. In \citep{Bhattacharyya:2007lq} this intuition is shown to be correct, provided the functions $b(x^\mu)$ and $\beta_i(x^\mu)$ obey a set of equations of motion, which turn out simply to be the equations of
boundary fluid dynamics.

Einstein's equations, when evaluated on the metric $G^{(0)}$, yield terms
which involve derivatives of the temperature and velocity fields in the boundary directions 
(i.e., $(x_i, v) \equiv x^\mu$) which we can organise order by order in a gradient expansion. 
Note that since $G^{(0)}$ is an exact solution to Einstein's equations when these fields are constants, terms with no derivatives are absent from this expansion. It is then  possible to show that field theory derivatives of either $\ln b(x^\mu)$ or $\beta_i(x^\mu)$ always
appear together with a factor of $b$. As a result, the contribution of $n$ derivative 
terms to the Einstein's equations is suppressed (relative to
terms with no derivatives) by a factor of $(b/L)^n \sim 1/(T\, L)^n$. Here
$L$ is the length scale of variations of the temperature and velocity
fields in the neighbourhood of a particular point, and $T$ is the temperature
at that point. Therefore, provided $L\,T \gg 1$, it is sensible to solve Einstein's
equations perturbatively in the number of field theory derivatives.\footnote{Note that the  variation in the radial direction, $r$, is never slow. Although we work order by order in the field theory derivatives, we will always solve all differential equations in the $r$ direction exactly. This should be contrasted with the holographic renormalization group which is a perturbative expansion in the Fefferman-Graham radial coordinate \cite{deHaro:2000xn}.} Essentially we are requiring that 
\begin{equation}
\frac{\p u^\mu}{T} \ , \frac{\p \log T}{T} \sim \ord{\varepsilon} \ll 1
\label{}
\end{equation}	
where we introduce a book-keeping parameter $\varepsilon$ which keeps track of derivatives in the boundary directions. It is useful to regard $b$ and $\beta_i$ as functions of the rescaled field theory coordinates $\varepsilon \, x^\mu$ where
$\varepsilon$ is a formal parameter which we eventually  set to unity. This way every derivative of $\beta_i$ or $b$ produces a power of $\varepsilon$, consequently powers of $\varepsilon$ count the number of derivatives.

\subsection{The perturbative expansion in gravity}
\label{s:pertexp}

We now describe a procedure to solve Einstein's equations in
a power series in $\varepsilon$. Consider the metric\footnote
{For convenience of notation we are dropping the spacetime indices in 
$G^{(n)}$. We also suppress the dependence of $b$ and $\beta_i$ on $x^\mu$.}
\begin{equation}\label{perturbansatz}
G = G^{(0)}(\beta_i, b) + \varepsilon \,G^{(1)}(\beta_i, b) + \varepsilon^2 \, G^{(2)}(\beta_i, b) + \CO\left(\varepsilon^3\right) ,
\end{equation}
where $G^{(0)}$ is the metric \eqref{boostedg0} and $G^{(1)}, G^{(2)}$ etc., are correction metrics that are yet to be determined.  We will see that  perturbative solutions to the gravitational equations exist only when the velocity and temperature fields obey certain equations of motion. These equations are corrected order by order in the $\varepsilon$ expansion; this forces us to correct the velocity and temperature fields themselves, order by order in this expansion. Consequently we set
\begin{equation} \label{veltempexp}
\beta_i= \beta_i^{(0)}+ \varepsilon \, \beta_i^{(1)}+ \CO\left(\varepsilon^2\right), \qquad b=\bz+ \varepsilon \, \bo + \CO\left(\varepsilon^2\right) , 
\end{equation}
where $\beta_i^{(m)}$ and $b^{(n)}$ are all functions of $\varepsilon \,x^\mu$.

In order to proceed with the calculation, it will be useful to fix a gauge.
In \citep{Bhattacharyya:2007lq} it was convenient to  work with the `background field'
gauge
\begin{equation}
G_{rr}= 0 \ , \qquad G_{r \mu }\propto u_\mu \ , \qquad {\rm Tr}\left(
(G^{(0)})^{-1} G^{(n)} \right)=0 \;\;\; \forall \; n  > 0 . 
 \label{gaugecond}
\end{equation}
One could equivalently work with a slightly different gauge choice \citep{Bhattacharyya:2008mz}:
\begin{equation}
G_{rr}= 0 \ , \qquad G_{r \mu } =  u_\mu \ . 
 \label{gaugecond2}
\end{equation}
With this  gauge choice it transpires that curves of $x^\mu = \text{constant}$ are in fact affinely parameterized null geodesics in the resulting spacetime, with the radial coordinate $r$ being the affine parameter. With the former gauge choice \req{gaugecond} of course $x^\mu = \text{constant}$ are null geodesics; however, $r$ is not an affine parameter for this geodesic congruence.

With this picture in place one can plug in the ansatz \req{perturbansatz} and \req{veltempexp} into Einstein's equations \req{ein} and expand them order by order in $\varepsilon$.  Let us imagine that we have solved the perturbation theory to the
$(n-1)^{{\rm th}}$ order, i.e., we have determined $G^{(m)}$ for $m\leq n-1$, and
have determined the functions $\beta_i^{(m)}$ and $b^{(m)}$ for $m \leq n-2$.
Plugging the expansion \eqref{perturbansatz} into Einstein's equations,
and extracting the coefficient of $\varepsilon^n$,  we
obtain an equation of the form
\begin{equation} \label{homoop}
{\mathbb H}\left[G^{(0)}(\beta^{(0)}_i, \bz)\right] G^{(n)}(x^\mu ) = s_n  .
\end{equation}
Here ${\mathbb H}$ is a linear differential operator of second order
in the variable $r$ alone. As $G^{(n)}$ is already of order $\varepsilon^n$,
and since every boundary derivative appears with an additional power of
$\varepsilon$, ${\mathbb H}$ is an ultralocal operator in the field theory directions.
In fact not only is  ${\mathbb H}$ is a differential operator only in the variable $r$ independent of $x^\mu$, but also its  precise form at the point $x^\mu$ depends only on the values of $\beta^{(0)}_i$ and $\bz$ at $x^\mu$, and not on the derivatives of these functions at that point. Furthermore,  the operator ${\mathbb H}$ is independent of $n$; we have the same homogeneous operator at every order in perturbation theory. This makes the perturbation expansion in $\varepsilon$  ultra-local in the boundary directions; we can solve the equations point by point on the boundary!

The source term $s_n$  however is different at different orders in perturbation
theory. It is a local expression of $n^{{\rm th}}$ order in boundary derivatives
of $\beta^{(0)}_i$ and $\bz$, as well as of $(n-k)^{{\rm th}}$ order in $\beta_i^{(k)}$, $b^{(k)}$ for all $k \leq n-1$. Note that $\beta_i^{(n)}$ and $b^{(n)}$ do not enter the $n^{{\rm th}}$ order equations as constant (derivative free) shifts of velocities
and temperatures solve the Einstein's equations.

The gravitational equation \eqref{homoop} form a set of $\frac{(d+1)(d+2)}{2}$ equations. It is useful to split these into two classes of equations: (i) a class that determines the metric data we need, comprising of $\frac{d(d+1)}{2}$ equations which we view as {\em dynamical equations} and (ii) a second set of $d$ equations which are essentially {\em constraint equations}.

\paragraph{Constraint equations:} We will refer to those of the Einstein's equations that are of first order in $r$ derivatives as constraint equations. These are obtained by contracting the equations with the one-form normal to the boundary
\begin{equation}
E^{(c)}_M = E_{MN} \, \xi^N
\label{}
\end{equation}	
where for our considerations $\xi_N = dr$. Of these equations, those with legs along the boundary direction are simply the equations of boundary energy momentum 
conservation:
\begin{equation} 
\nabla_\mu T_{(n-1)}^ {\;\mu \nu}=0 \ .
\label{stcons} 
\end{equation}
Here $T_{(n-1)}^ {\;\mu \nu} $ is the boundary stress tensor dual the solution
expanded up to $\CO\left(\varepsilon^{n-1}\right)$ and is a local function of the temperature and velocity fields involving no more than $n-1$ derivatives. Furthermore, it is conformally covariant and consequently it is a fluid dynamical stress tensor with $n-1$ derivatives.

These constraint equations can be used to determine $b^{(n-1)}$ and $\beta_i^{(n-1)}$; this is essentially solving the fluid dynamics equations at order $\ord{\varepsilon^n}$ in the gradient expansion assuming that the solutions at preceding orders are known. There is a 
non-uniqueness in these solutions given by the zero modes obtained by  linearizing the equations of stress energy conservation at zeroth order. These can be absorbed into a redefinition of the  $\beta^{(0)}_i, b^{(0)}$, and do not correspond to a physical non-uniqueness. 

\paragraph{Dynamical equations:}  The remaining constraint $E_{rr}$ and the dynamical Einstein's equations  $E_{\mu\nu}$ can then be used to solve for the unknown function $g^{(n)}$. By exploiting the underlying symmetries of the zeroth order solution, specifically the rotational symmetry in the spatial sections on the boundary, $SO(d-1)$, it is possible to decouple the system of equations into a set of first order differential operators. 
Having performed this diagonalization of the system of equations one has a formal solution of the form:
\begin{equation}
G^{(n)} = {\rm particular}(s_n) + {\rm homogeneous}({\mathbb H})
\label{gnexpr}
\end{equation}	
To determine the solution uniquely we need to prescribe boundary conditions: we impose that our solution is normalizable so that the spacetime is asymptotically \AdS{d+1} and also demand regularity at all $r \neq 0$. In particular, the solution should be regular at the hypersurface $b\,r =1$. It has been shown  in \citep{Bhattacharyya:2007lq} that for an arbitrary  non-singular and appropriately normalizable source $s_n$  encountered in   perturbation theory, it is always possible to choose these constants to ensure that $G^{(n)}$ is appropriately normalizable  at $r=\infty$  and non-singular at all nonzero $r$. Furthermore, if the solution
at order $n-1$ is non-singular at all nonzero $r$, it is guaranteed to produce a 
non-singular source at all nonzero $r$. Consequently, the non-singularlity of 
$s_n$ follows inductively.\footnote{There is a slight subtlety which needs to be borne in mind here: the requirements above do not completely fix $G^{(n)}$ since the differential operator ${\mathbb H}$ has some zero modes. These can however be fixed by appropriately absorbing the zero modes into redefinition of the local temperature and velocity fields.}

\subsection{Details of the long-wavelength perturbation expansion}
\label{s:details}

We have described how to perturbatively construct  solutions to \req{ein} by working order by order in boundary derivatives. We now briefly illustrate how to carry this out in practice to first order in derivatives and refer the reader to the original references \citep{Bhattacharyya:2007lq,Bhattacharyya:2008mz} for the detailed derivation of the results given herein.

Consider the zeroth order metric $G^{(0)}$ given in \req{boostedg0}. If we want to work to first order in boundary derivatives, we can pick a point on the boundary $x^\mu = x^\mu_0$, which by exploiting the Killing symmetries of the background can be chosen to be the origin. At $x^\mu_0$ we can use the local scaling symmetry to set $b^{(0)} = 1$ and pass to a local inertial frame so that $\beta^{(0)}_i=0$. Expanding \req{boostedg0} to  $\ord{\varepsilon}$ at this point we find
\begin{equation}
\begin{split}
ds_{(0)}^2 &= 2\, dv \, dr -r^2\, f(r)\, dv^2 + r^2\, dx_i \, dx^i \\
&- 2\, \delta\beta^{(0)}_i \, dx^i \, dr -
2\, \delta\beta^{(0)}_i \, r^2\, (1-f(r))\, dx^i \, dv  - d \,\frac{ \delta \bz}{r^{d-2}}\, dv^2 \  , 
\end{split}
\label{fote}
\end{equation}	
where we have introduced a short-hand
 $\delta \beta^{(0)}_i = x^\mu \,\p_\mu \beta^{(0)}_i$ and $\delta b^{(0)} = x^\mu\,\p_\mu b^{(0)}$ which are the leading terms in the Taylor expansion of the velocity and temperature fields at $x_0^\mu$ (taken to be the origin).
 
 We now need to pick an ansatz for the metric correction at $\ord{\varepsilon}$, $G^{(1)}$, which we wish to determine. As mentioned earlier it is useful to exploit the $SO(d-1)$ spatial rotation symmetry at $x^\mu_0$ to decompose modes into various representations of this symmetry.  Modes of $G^{(1)}$ transforming under different representations decouple from each other by symmetry. We have  the following decomposition into $SO(d-1)$ irreps:
\begin{equation}
\begin{aligned}
& \text{scalars:} &\qquad & G^{(1)}_{vv},  G^{(1)}_{vr}, \sum_i\,  G^{(1)}_{ii},  \\
&\text{vectors:} & \qquad & G^{(1)}_{vi}, \\
&\text{tensors:} &\qquad & G^{(1)}_{ij}
\end{aligned}
\label{grpdecomp}
\end{equation}	
We work sector by sector and solve the constraint and the dynamical Einstein's equations.

In the scalar sector, we find that the constraint equations imply that 
\begin{equation}
\frac{1}{d-1}\, \p_i\beta^{(0)}_i =  \p_v b^{(0)}  \ , 
\label{scalarcons}
\end{equation}	
 while in the vector sector we have
\begin{equation}
 \p_i b^{(0)} =  \p_v \beta^{(0)}_i \ .
\label{veccons}
\end{equation}	
These two equations are simply the equations of energy momentum conservation \req{cons1} at the point $x^\mu_0$. The remaining dynamical equations are to be solved for the functions appearing in $G^{(1)}$ -- we refer the reader to \citep{Bhattacharyya:2007lq,Bhattacharyya:2008mz} for the specifics and just record here the form of the differential operator we obtain in various sectors:\footnote{We have indexed the operator ${\mathbb H}$ by the representation label of the $SO(d-1)$ rotational symmetry. The scalar sector involves some mixing between different fields and is slightly more involved, see \citep{Bhattacharyya:2007lq,Bhattacharyya:2008mz} for details.}
\begin{eqnarray}
\text{vector}:&& {\mathbb H}_{{\bf d-1}} \CO = \frac{d}{dr}\,\left(\frac{1}{r^{d-1}}\, \frac{d}{dr}\,\CO\right) \nonumber \\
\text{tensor}:&& {\mathbb H}_{{\bf \frac{d(d+1)}{2}}} \CO = \frac{d}{dr}\,\left(r^{d+1}\, f(r)\, \frac{d}{dr}\,\CO\right)
\label{Hvar}
\end{eqnarray}	
which as advertised earlier are simple differential operators in the radial variable alone and clearly can be inverted to find the function $\CO$ once the source $s_n$ is specified. 

The calculation can in principle be carried out to any desired order in the $\varepsilon$ expansion. As we have discussed earlier the form of the differential operator \req{Hvar} remains invariant in the course of the perturbation expansion. What one needs to compute at any given order are the source terms $s_n$. In addition one has to always ensure that the lower order stress tensor conservation equations are satisfied. For instance in order for the source terms which appear in the determination of $G^{(2)}$ at second order to be ultra-local  at our chosen boundary point $x^\mu_0$ we have to ensure that the first order fluid equations of motion are satisfied. This is encapsulated in our discussion of the constraint equations in the bulk \req{stcons}.

\section{Gravitational analysis: Metrics dual to fluids}
\label{s:grav2}

We have thus far discussed how to solve to Einstein's equations order by order in boundary derivatives. We now present the result for the general fluid dynamical metric up to second order in boundary derivatives and then describe how one extracts the stress tensor quoted in \sec{s:nln4}. Subsequently, we will analyze the physical properties of these solutions and argue that they are regular black hole solutions with an event horizon. Following that we will discuss how one can use the black hole nature of the solution to understand aspects of the fluid dynamics such as the entropy current. 

\subsection{The gravitational dual to non-linear viscous fluid}
\label{s:gravnl}

We have described how one can consistently solve for the bulk metric in \sec{s:gravity} -- in particular, in \sec{s:details} we have given a sketch of  how the perturbation scheme works. It turns out that the bulk metric  resulting from the explicit computation can schematically be cast into the form:
\begin{equation}
ds^2= G_{MN} \,dX^M \,dX^N = - 2 \, \CS(r,x)\,  u_\mu(x) \, dx^\mu \, dr+ \chi_{\mu \nu}(r,x) \, dx^\mu\, dx^\nu  \  .
\label{formmet}
\end{equation}
This was the form of the metric originally derived in \citep{Bhattacharyya:2007lq}, where the functions $\CS(r,x)$ and $\chi(r,x)$ are explicit functions of the radial coordinate $r$, whilst being given in terms of a gradient expansion in the boundary directions $x^\mu$.  Note that the bulk metric is actually given in the gauge \req{gaugecond}. The expressions for the functions $\CS(r,x)$ and $\chi(r,x)$ are rather cumbersome and  we will not record them here, but rather refer the interested reader to the original source \citep{Bhattacharyya:2007lq,Bhattacharyya:2008xc}. Instead we will record the explicit metric below making manifest the Weyl covariant structures, following \citep{Bhattacharyya:2008mz}  which makes for a more compact presentation.  Before we do that however, it is useful to understand the geometry in the somewhat simpler form of the metric given above, since it captures much of the essential physical features.

First of all, it is useful to realize that in the metric \req{formmet} lines of constant $x^\mu$ are radially ingoing null geodesics. If the function $\CS(r,x^\mu)$ is gauge fixed to unity then in fact the radial coordinate $r$ is actually an affine parameter along this null congruence. One can visualize this as follows: consider a null geodesic congruence i.e., a family of null geodesics with exactly one geodesic passing through each point, in some region of an arbitrary spacetime. Let $\Sigma$ be a hypersurface
that intersects each geodesic once. Let $x^\mu$ be coordinates on
$\Sigma$. Now ascribe coordinates $(\rho,x^\mu)$ to the point at an 
affine parameter distance $\rho$ from $\Sigma$, along the geodesic through the
point on $\Sigma$ with coordinates $x^\mu$. Hence the geodesics in the
congruence are lines of constant $x^\mu$. In this chart, this metric
takes the form 
\begin{equation}
 ds^2 = -2 \,u_\mu(x) \,d\rho \,dx^\mu +
\widehat{\chi}_{\mu\nu}(\rho,x)\, dx^\mu \,dx^\nu, 
\end{equation}
where the geodesic equation implies that $u_\mu$ is independent of $\rho$.  It is
convenient to generalize slightly to allow for non-affine
parametrization of the geodesics: let $r$ be a parameter related to
$\rho$ by $d\rho/dr = {\cal S}(r,x)$. Then, in coordinates
$(r,x)$, the metric takes the form given in \req{formmet}. Note that $\Sigma$ could be spacelike, timelike, or null. We shall take $\Sigma$
to be timelike. 

The metric \req{formmet} has determinant $-{\cal S}^2 \chi^{\mu\nu}
u_\mu u_\nu \det \chi$, where $\chi^{\mu\nu}$ is the inverse of
$\chi_{\mu\nu}$, where the indices are raised with the boundary metric; for \req{formmet} the induced metric on the boundary $\CB_d$ is simply 
$$g_{\mu\nu} = \lim_{r\to \infty}  \; \frac{1}{r^2} \, \chi_{\mu\nu}(r,x)\ .$$ 
Hence the metric and its inverse will be smooth if
${\cal S}$, $u_\mu$ and $\chi_{\mu\nu}$ are smooth, with ${\cal S}
\ne 0$, $\chi_{\mu\nu}$ invertible, and $\chi^{\mu \nu} \, u_\mu$ timelike. These conditions are satisfied on, and outside, the horizons of the solutions that we shall discuss below. Finally, note that the inverse metric,\footnote{Note that the `inverse $d$-metric' $\chi^{\mu\nu}$ is defined via $\chi_{\mu\nu} \, \chi^{\nu\rho} = \delta_\mu^{\ \rho}$.} $G^{MN}$, can be determined easily to be 
\begin{equation}
G^{rr} = \frac{1}{-\CS^2 \,  u_\mu \, u_\nu \, \chi^{\mu\nu}} \ , \qquad
G^{r\alpha} = \frac{\CS \, \chi^{\alpha\beta}\,u_{\beta}}{- \CS^2 \, u_\mu \, u_\nu \, \chi^{\mu\nu}} \ , \qquad
G^{\alpha\beta} = \frac{\CS^2 \, u_ \gamma \, u_ \delta \, \left(
\chi^{\alpha\beta} \,\chi^{\gamma\delta} - \chi^{\alpha \gamma} \,\chi^{\beta \delta}  \right)}{- \CS^2 \, u_\mu \, u_\nu \, \chi^{\mu\nu}} \ ,
\label{invmet}
\end{equation}	
using  $G_{MK}\,G^{KN} = \delta_M^{\ N}$.

\paragraph{Weyl covariant form of fluid metric:} 
As remarked above the coordinates in which we present the fluid metric \req{formmet}  do not make explicit the Weyl covariant structures. An alternate form of the metric was written down in general for asymptotically \AdS{d+1} spacetimes in \citep{Bhattacharyya:2008mz} which takes the form:
\begin{equation} 
ds^2 =G_{MN} \,dX^M \,dX^N = - 2 \,   u_\mu(x) \, dx^\mu \,\left( dr + \bCV_\nu(r,x)\,\,dx^\nu\right)+ \bCG_{\mu \nu}(r,x) \, dx^\mu\, dx^\nu \ , 
\label{formmetw}
\end{equation}
where the fields $\bCV_\mu$ and $\bCG_{\mu\nu}$ are functions of $r$ and $x^\mu$ which admit an expansion in the boundary derivatives. In the parameterization used in \citep{Bhattacharyya:2008mz} one finds the metric functions are given up to second order in derivatives as:
\begin{equation}
\begin{aligned}
\bCV_\mu & = r\, \CA_\mu - \CS_{\mu\lambda}\,u^\lambda - \bcv_1(b\,r)\, P^{\; \nu}_\mu\, \WD_\lambda\sigma^{\lambda}_{\ \nu} \\
&\qquad  +u_\mu \, \left[\frac{1}{2}\, r^2 \, f( b\,r) + \frac{1}{4}\, \left(1-f(b\,r)\right) \, \omega_{\alpha\beta}\, \omega^{\alpha\beta} + \bcv_2(b\,r) \, \frac{\sigma_{\alpha\beta}\, \sigma^{\alpha\beta}}{d-1}\right]  \\
\bCG_{\mu\nu} &= r^2\, P_{\mu\nu}  - \omega_{\mu}^{\ \lambda}\,\omega_{\lambda\nu}+ 2\, (b\,r)^2\, \bcg_1(b\,r)\, \left[\frac{1}{b}\, \sigma_{\mu\nu} + \bcg_1(b\,r) \, \sigma_\mu^{\ \lambda}\,\sigma_{\lambda\nu} \right]- \bcg_2(b\,r)  \,\frac{\sigma_{\alpha\beta}\, \sigma^{\alpha\beta}}{d-1} \, P_{\mu\nu} \\
& \qquad \,- \bcg_3(b\,r) \, \left[\bCT_{1\mu\nu} + \frac{1}{2}\, \bCT_{3\mu\nu}   + 2\,\bCT_{2\mu\nu} \right] + \bcg_4(b\,r)\,  \left[\bCT_{1\mu\nu} +  \bCT_{4\mu\nu}   \right] .
\end{aligned}
\label{met2w}
\end{equation}
where we are using the tensors defined earlier in \req{winv2der} and also introduce the Schouten tensor $\CS_{\mu\nu}$ (a particular combination of curvature tensors) which is defined in \req{schouten:eq}. The tensor $\bCG_{\mu\nu}$ is clearly transverse, since it is built out of operators that are orthogonal to the velocity, and it can be inverted via the relation 
$\left(\bCG^{-1}\right)^{\mu\alpha} \, \bCG_{\alpha\nu}  = P^\mu_{\;\nu}$.
 We also note that the coordinate chart used to write \req{formmet} is consistent with the bulk gauge choice \req{gaugecond2}. For  future reference we record the induced metric on the boundary in these coordinates
\begin{equation}
g_{\mu\nu} = \lim_{r\to \infty} \, \frac{1}{r^2} \, \left(\bCG_{\mu\nu} - u_{(\mu}\, \bCV_{\nu)} \right) ,
\label{bdymet2w}
\end{equation}	
which is crucially used to raise and lower the boundary indices (lowercase Greek indices). 

The various functions appearing in the metric are given in terms of definite integrals 
\begin{equation}
\begin{aligned}
\bcg_1(y) &= \int_y^\infty\, d\zeta\, \frac{\zeta^{d-1}-1}{\zeta\, \left(\zeta^d -1\right)} \\
\bcg_2(y) &= 2\,y^2 \, \int_y^\infty \frac{d\xi}{\xi^2} \int_\xi^\infty\, d\zeta\,\zeta^2 \, \bcg_1'(\zeta)^2 \\
\bcg_3(y) & =y^2 \, \int_y^\infty\, d\xi\, \frac{\xi^{d-2}-1}{\xi \, \left(\xi^d -1\right)} \\
\bcg_4(y)& = y^2 \, \int_y^\infty\,  \frac{d\xi}{\xi \, \left(\xi^d -1\right)}
\int_1^\xi\,d\zeta\, \zeta^{d-3}\bigg(1+ (d-1)\,\zeta\,\bcg_1(\zeta) + 2\, \zeta^2\,\bcg'_1(\zeta)\bigg) \\
\bcv_1(y) &=\frac{2}{y^{d-2}} \, \int_y^\infty \, d\xi \; \xi^{d-1}\int_\xi^\infty\, d\zeta\, \frac{\zeta-1}{\zeta^3\, \left(\zeta^d-1\right)}\\ 
\bcv_2(y) &= \frac{1}{2\, y^{d-2}} \, \int_y^\infty\; \frac{d\xi}{\xi^2}\,\bigg[1-\xi\,(\xi-1) \,\bcg'_1(\xi)  -2 \,(d-1)\,\xi^{d-1} \\
&\qquad + \left(2\, (d-1)\,\xi^d - (d-2)\right) \, \int_\xi^\infty\, d\zeta\, \zeta^2\,\bcg'_1(\zeta)^2 \bigg]  .
\end{aligned}
\label{fmetfnsw}
\end{equation}
The asymptotic behaviour of these functions $\bcg_i(y)$ and $\bcv_i(y)$  which are relevant for the stress tensor computation can be found in \citep{Bhattacharyya:2008mz}.

\subsection{The boundary stress tensor}
\label{s:stensor}

Once the bulk black hole solution is determined it is straightforward to use the holographic prescription of \citep{Henningson:1998gx,Balasubramanian:1999re} to compute the boundary stress tensor. To perform the computation we regulate the asymptotically \AdS{d+1} spacetime at some cut-off hypersurface $r  = \Lambda_\text{c}$ and consider the induced metric on this surface, which up to a scale factor involving $\Lambda_{\text{c}}$ is our boundary metric $g_{\mu\nu}$. The holographic stress tensor is given in terms of the extrinsic curvature $K_{\mu\nu}$ and metric data of this  cut-off hypersurface. Denoting the unit outward normal to the surface by $n^\mu$ we have 
\begin{equation}
K_{\mu\nu} = g_{\mu\rho}\,\nabla^\rho n_\nu
\label{extrinsicc}
\end{equation}	
For example for asymptotically \AdS{5} spacetimes the prescription of \citep{Balasubramanian:1999re} gives
\begin{equation}
T^{\mu\nu} = \lim_{\Lambda_\text{c} \to \infty}\; \frac{\Lambda_\text{c}^{d-2}}{16\pi \, G_N^{(d+1)}} \, \left[ K^{\mu\nu} - K \, g^{\mu\nu} - (d-1)\, g^{\mu\nu} - \frac{1}{d-2}\,  \left(R^{\mu\nu} -\frac{1}{2}\, R \, g^{\mu\nu}\right)\right] 
\label{}
\end{equation}	
where $K^{\mu\nu}$ is the extrinsic curvature of the boundary. Implementing this procedure for the metric \req{formmetw} and utilizing the asymptotic form of the functions given in \req{fmetfnsw} we recover the stress tensor quoted in \req{fld2} with the precise transport coefficients \req{transpgend}.

\subsection{Event horizons}
\label{s:eventhor}
Having understood the geometric aspects of the coordinate chart employed for metrics dual to arbitrary fluid flows in the boundary, we are now in a position to address our assertion that these metrics are in fact black hole spacetimes with regular event horizon. In general determining the event horizon of a black hole spacetime requires that we know the entire future evolution of the spacetime geometry. This is basically due to the definition of an event horizon, which is teleological in nature. 

Formally one defines the event horizon of a given spacetime as follows: the future event horizon $\CH^+$ is the boundary of past set of future null infinity. This is the mathematical way to capture the physical statement that the spacetime events inside the event horizon of the black hole cannot communicate to the asymptotic region. It is important to note that the future null infinity $\scri^+$ corresponds to where the future directed null geodesics in the spacetime end and is in fact timelike for asymptotically AdS spacetimes. Since $\CH^+$ is the boundary of a causal set, it is a null surface which is in particular generated by null geodesics in the spacetime. One can determine the event horizon by shooting null geodesics from the interior of the spacetime and checking whether they make it out to $\scri^+$. By fine tuning the initial conditions of the geodesic one can zero onto $\CH^+$. Alternately, if one knows the late time generators of the event horizon, then one can evolve the geodesic equation backwards with these generators as the future boundary condition. 

For the metrics dual to boundary fluid dynamics, we will make an important assumption: since fluid dynamical evolution tends generically to smooth out inhomogeneities, we will assume that the late time solution is one corresponding to global equilibrium wherein the fluid settles down. Then at late times we have a clear idea of where the event horizon is located -- its location in the radial direction in the asymptotically \AdS{d+1} bulk spacetime is specified by the value of the local temperature. Armed with this data one can in principle evolve the null generators of the horizon backwards to construct the entire surface $\CH^+$.\footnote{As has been recently discussed in the context of dynamical fluid flows \citep{Figueras:2009iu} there are examples such as the the Bjorken flow and the conformal soliton flow where despite the long wavelength approximation required for fluid dynamics to be valid being upheld, one nevertheless has non-trivial late time boundary conditions which change the nature of the event horizon.} However, it will turn out that in the long-wavelength approximation one can as usual work order by order in the boundary derivative expansion and in fact determine the location of the horizon. 

For the spacetime \req{formmet}, let us suppose that the null hypersurface corresponding to $\CH^+$ that we are after is given by  the equation
\begin{equation} \label{ehsurf} 
S_{\CH}(r, x) =0 \ , \qquad {\rm with} \qquad 
S_{\CH}(r,x) =r-r_H(x) \ .
\end{equation}
As we are working in a derivative expansion and assuming that the dissipative fluid dynamics drives us towards global equilibrium with a local temperature $T(x) = \frac{d}{4\pi\, b(x)}$ we take
\begin{equation}
r_H(x)=  \frac{1}{b(x)}+ \sum_{k=1}^\infty \, \varepsilon^k \,r_{(k)}(x) \ .
\label{rhpert}
\end{equation}	
where $r_{(k)}(x)$ denote the corrections away from the late time hypersurface in the spacetime. By demanding that the hypersurface defined by \req{ehsurf} be null i.e.,
\begin{equation}\label{logic}
G^{AB}\, \partial_A S_{\CH} \, \partial_B S_{\CH} =0
\end{equation}
we can solve for the functions $r_{(k)}(x)$ order by order in the $\varepsilon$ expansion. In addition to the location of the horizon we will also be interested in the  normal vector to the event horizon by $\xi^A$: by definition,
\begin{equation}
\xi^A = G^{AB} \, \partial_B S_{\CH}(r, x)
\label{nadef}
\end{equation}	
which also has an $\varepsilon$ expansion. Using the explicit form of the metric \req{formmet} or \req{formmetw} one can  easily write down the equation  determining the location of the event horizon  \req{logic} to arbitrary order in $\varepsilon$  explicitly. For instance in the simpler situation of the non-Weyl covariant metric \req{formmet} we have the equation
\begin{equation}
0 =  \frac{1}{-\CS^2 \,  u_\mu \, u_\nu \, \chi^{\mu\nu}} 
\left( 1 - 2 \,\varepsilon \, \CS \, \chi^{\alpha\beta}\,u_{\beta} \, \partial_\alpha r_H 
- \varepsilon^2 \, \CS^2 \, \left(
\chi^{\alpha\beta} \,\chi^{\gamma\delta} - \chi^{\alpha \gamma} \,\chi^{\beta \delta}  \right) 
\, u_ \gamma \, u_ \delta \, \partial_ \alpha r_H \, \partial_ \beta r_H 
\right) , 
\label{horizon}
\end{equation}	
which is actually an algebraic equation for the functions $r_{(k)}$ at any given order in $\varepsilon$ and hence easy to solve. For the Weyl covariant metric \req{formmetw}, the horizon location is determined by the equation \citep{Bhattacharyya:2008mz}:
\begin{equation}
\left(\bCG^{-1}\right)^{\mu\nu} \, \kappa_\mu\,\kappa_\nu = 2\, u^\mu\, \kappa_\mu \ , 
\label{horizonw}
\end{equation}	
with 
\begin{equation}
\kappa_\mu = \partial_\mu r_H + \bCV_\mu(r_H(x),x) \  , 
\label{kappadef}
\end{equation}	
where we have left the powers of $\varepsilon$ implicit. In terms of $\kappa_\mu$ one can determine the components of the normal vector to the horizon hypersurface as 
\begin{equation}
\xi^A\,\p_A = n^\mu\left(\p_\mu + \p_\mu r_H\, \p_r\right) , \;\; \text{with} \;\; \;n^\mu = u^\mu - \left(\bCG^{-1}\right)^{\mu\nu} \, \kappa_\nu  \ . 
\label{normalw}
\end{equation}	

Rather than illustrate the computation of the event horizon for the metrics \req{formmetw} or \req{formmetw} we will just give the resulting answer of the computation, referring the reader to the original derivation presented in \citep{Bhattacharyya:2008xc,Bhattacharyya:2008mz}. However, before doing so we would like to present a toy model computation which illustrates the key features of the computation.

\paragraph{Toy model for event horizon detection:} The general situation we are interested in has dependence on all the boundary directions $x^\mu$ -- a simpler situation to consider would be to look at a case where we have dependence only on one variable, say time. Furthermore, there is nothing really special about asymptotically AdS spacetimes for the detection of the event horizon. One could just as readily have made similar arguments for asymptotically flat spacetimes. As a result we will focus on a simple example of the Vaidya spacetime which describes a spherically symmetric black hole with infalling null matter. As we are interested in just the geometric properties of the spacetime, we will just focus on the metric and not worry about solving Einstein's equations.

Consider the Vaidya spacetime, whose metric is given as:
\begin{equation}
 ds^2 = - \left( 1 - \frac{2\,m(v)}{r} \right) \,dv^2 + 2\, dv \,dr + r^2 \,d\Omega^2 \ .
\label{vaidya}
\end{equation}
We want to determine the location of the horizon, which by spherical symmetry has to lie on the locus  $r=r(v)$. The normal one-form to this surface is clearly just $n = dr - \dot{r} \, dv$. Demanding that this be null gives us a differential equation for the null surface 
\begin{equation}
r(v) = 2\, m(v) + 2\, r(v)\,  \dot{r}(v) \ ,
\label{vode}
\end{equation}	
which is the analog of \req{horizon} for the Vaidya metric \req{vaidya}. The equation we have at hand is a first order ODE for the function $r(v)$, solving which we would determine the location of the horizon  {\it non-locally} in terms of $m(v)$. This would of course be the case for a generic function $m(v)$. 

 However, to make contact with our earlier discussion, suppose we assume that the mass function $m(v)$ is slowly varying and moreover that it asymptotes to a constant for late times, i.e., 
\begin{equation}
\dot{m}(v)=\ord{\varepsilon}  \ , m \,  \ddot{m} = \ord{\varepsilon^2}  \; \text{etc.}, \qquad {\rm  and} \qquad  \lim_{v \to \infty} m(v) = m_0 
\label{}
\end{equation}	
then we can solve \req{vode} in a derivative expansion. Consider the ansatz, 
\begin{equation}
r(v) = 2\, m(v) + \sum_{k=1}^\infty \, \varepsilon^k\, r_{(k)}(v) 
\label{}
\end{equation}	
which leads to the solution
\begin{equation}
r_{(1)}= 8\, m \, \dot{m}  \ , \qquad r_{(2)}= 64\, m \, \dot{m}^2 + 32\, \ddot{m} \, m^2 , \quad  \text{etc}..
\label{}
\end{equation}	
Thus,  by invoking the slow variation  ansatz we  obtain a {\it local} expression for the
location of the horizon in a derivative expansion. 

\begin{figure}[h!]
\begin{center}
\includegraphics[scale =0.8]{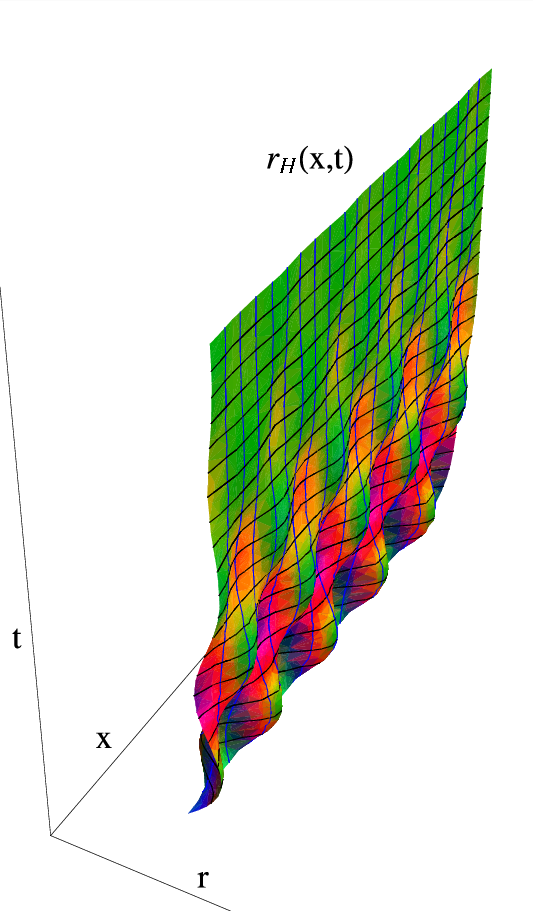}
\end{center}
\caption{The event horizon $r=r_H(x^\mu)$ sketched as a function of the time $t$ and one of the spatial coordinates $x$ (the other two spatial coordinates are suppressed).} 
\label{wigglyhor}
\end{figure}

\paragraph{Horizon location for fluid metrics:} An analogous analysis can be carried out for the the equation \req{horizonw} -- we quote here just the final result (once again dropping $\varepsilon$ for brevity)
\begin{equation}
r_H(x) = \frac{1}{b(x)} + b(x) \, \left(\aleph_1 \, \sigma_{\alpha\beta}\, \sigma^{\alpha \beta} + \aleph_2 \, \omega_{\alpha\beta}\, \omega^{\alpha\beta} + \aleph_3\, \CR\right)
\label{}
\end{equation}	
with
\begin{equation}
\begin{aligned}
\aleph_1 &= \frac{2\,(d^2 +d-4)}{d^2\,(d-1)\,(d-2)} - \frac{2 \,\bcv_2(1)}{d\,(d-1)}\\
\aleph_2 &= -\frac{d+2}{2\,d\,(d-2)} \\ \aleph_3 &= -\frac{1}{d\,(d-1)\,(d-2)}
\end{aligned}
\label{alephdef}
\end{equation}	
$\CR$ here is the Weyl covariant boundary Ricci scalar, which is defined in \req{ricEin:eq}.
This establishes the claim made in the \sec{s:intro} that spacetimes dual to hydrodynamical evolution in the boundary field theory are regular black hole spacetimes. See \fig{wigglyhor} for an illustration of the event horizon in the spacetime.

\subsection{Boundary entropy current}
\label{s:entcur}

As we discussed in \sec{s:dissf} the flow of a dissipative fluid is characterized by entropy production. Given that we have constructed a gravitational dual which is a black hole it is natural to ask whether one can associate the entropy of the fluid with a geometric feature in the bulk spacetime. There is a natural object which one is tempted to use for this purpose, which is the area of the event horizon in the bulk geometry. 

In general when we consider deviations from equilibrium it is a-priori not clear that there should exist an unambiguous notion of entropy.  In fact, since all we require of the entropy current is to satisfy the second law \req{seclaw}, we can use any local function having positive divergence to characterize the irreversibility of the fluid dynamical flow. The only constraint we would demand on this putative Boltzmann H-function  is that it agrees with the thermodynamic notion of entropy in global equilibrium. For a stationary black hole, which corresponds to global thermal equilibrium,  the area of the event horizon does indeed capture the entropy of the dual field theory. It seems therefore natural to associate the entropy of the field theory with the area of the event horizon. This actually turns out to be a bit more subtle as discussed recently in \citep{Figueras:2009iu}-- we will return to this point after providing a sketch of the argument given in \citep{Bhattacharyya:2008xc} for constructing an entropy current dual to fluid flows using event horizon data.\footnote{The rationale behind postponing the discussion of the subtleties is that the event horizon serves to illustrate the  general idea  of defining a  boundary entropy current; it is trivial to generalize the construction to other quasi-local horizons.}

\paragraph{Entropy current from geometry:} We have established that the spacetimes dual to boundary fluid dynamics are in fact black hole geometries with a regular future event horizon $\CH^+$. Given a black hole spacetime one can associate an entropy with the area of the event horizon, which is the usual Bekenstein-Hawking entropy of the back hole. However, the field theory lives on the boundary of the spacetime and we would like to define an entropy current directly in the field theory. To this end, consider spatial sections of the event horizon, which are co-dimension two surfaces in the spacetime, which we label as $\CH^+_v$. We are working in a coordinate chart where the coordinates $\alpha^i$ for $i =\{1,\cdots d-1\}$ define a chart on  the spatial section and we use as affine parameter, the boundary coordinate  $v$, to propagate these surfaces forward along the horizon generator $\xi^A$.  On the surface $\CH^+_v$  it is natural to define an area $(d-1)$-form, whose integral gives the area of the spatial section. From the area theorem for black holes, it follows that the area of the spatial  sections will be non-decreasing with $v$.\footnote{We assume that the null energy condition is satisfied. This is clearly true of the lagrangian \req{ehact}, but we will demand the same to be true for the general discussion \req{ehactm}.} 

Once we have a geometric object such as the area-form on the horizon, we can pull it back to the boundary and have a candidate entropy function, for the pull-back too will have non-negative divergence.  The only issue is the precise manner in which we implement the pull-back. It turns out to be natural to pull-back the area form on the horizon using radially ingoing null geodesics, which provide an isomorphism between the spatial sections on the boundary  and the corresponding $\CH^+_v$ on $\CH^+$. This procedure was described in detail in \citep{Bhattacharyya:2008xc} and was used there to construct a boundary entropy current explicitly. One can give a compact formula for the boundary entropy current based on this discussion as follows. Let us suppose that the induced metric on the spatial slices of the horizon $\CH^+_v$ is given by ${\mathfrak h}_{ij}$ and as usual the boundary metric is $g_{\mu\nu}$. Under the split of the future horizons into spatial sections $\CH^+_v$ we can also split the vector $n^\mu$ defined in \req{normalw} which essentially gives us the projection of the null generator of the event horizon into the boundary directions. Armed with this set of geometric data we define an entropy current
\begin{equation}
J^\mu_S = \frac{1}{4\, G_N^{(d+1)}}\; \frac{\sqrt{\det_{d-1}{\mathfrak h}}}{n^v \, \sqrt{-\det{g}}} \, n^\mu
\label{entcdef}
\end{equation}	
which implements the pull-back procedure from the event horizon to the boundary.

\paragraph{Boundary entropy current:} The entropy current in $d$ dimensions has to be a Weyl covariant vector of weight $d$. This follows from the fact that the entropy density scales like the inverse spatial volume, the total entropy being dimensionless, and the scaling of the velocity field given by\req{velweyl}. In  \citep{Loganayagam:2008is}  the general constraints on entropy currents were analyzed in the Weyl covariant formalism and a particular entropy current for four dimensional CFTs was constructed. This analysis was slightly generalized in \citep{Bhattacharyya:2008xc} to argue for a five parameter family of fluid dynamical entropy currents in four dimensional CFTs. A similar analysis was carried out in arbitrary dimensions in \citep{Bhattacharyya:2008mz}, who found a four parameter family of hydrodynamical entropy currents. The extra parameter which appears in $d=4$ is related to the fact that one can have pseudo-vector contributions to the entropy current involving $\ell^\mu$ defined in \req{lmudef}. In general $d$ dimensions the entropy current takes the form:\footnote{The contribution from the pseudo-vector in four dimensions takes the form $C_1\,s\, b \,\ell^\mu + C_2 \,s\, b^2 \, u^\lambda \, \WD_\lambda \ell^\mu$. The positivity condition on the divergence of the entropy current demands $C_1 + C_2 =0$.}
\begin{equation}
\begin{split}
\, J^\mu_S &= s\,u^\mu + s\, b^2 \, u^\mu \,\left(A_1 \,\sigma_{\alpha\beta}\,\sigma^{\alpha\beta}+A_2 \,\omega_{\alpha\beta}\,\omega^{\alpha\beta} +A_3 \,\mathcal{R}\,\right)\\
&\quad  + s\, b^2 \,\left( B_1 \,\WD_\lambda \sigma^{\mu\lambda} + B_2 \,\WD_\lambda \omega^{\mu\lambda} \right) + \cdots
\end{split}
\label{ecurw}
\end{equation}
where $s$ is the entropy density and $A_{1,2,3}$, $B_{1,2}$ are arbitrary numerical  coefficients. Requiring positivity of the divergence one finds a single linear relation between two of the coefficients:
\begin{equation}
B_1 + 2\, A_3 = 0 \ . 
\label{ABeq}
\end{equation}	

The gravitational analysis described above leads to a specific entropy current, with the parameters fixed by the geometric data encoded in the metric \req{formmetw}. The essential data is captured the entropy density, which in general is given by the Bekenstein-Hawking formula in terms of the area of the event horizon\footnote{As discussed above in the situations where the temporal variation is suitably slow the ateleological behaviour of the event horizon doesn't play an important role. Furthermore, the various quasi-local horizons are `sufficiently near' the actual event horizon which makes it possible to use the area of the event horizon for our purposes.} 
\begin{equation}
s = \frac{1}{4\,G_N^{(d+1)}} \,\frac{1}{b^{d-1}} \  , \qquad b =  \frac{d}{4\pi\, T} \ . 
\label{sdengr}
\end{equation}	
The coefficients appearing in \req{ecurw} are given in terms of the numerical (dimension) dependent constants satisfying \req{ABeq}
\begin{equation}
\begin{split}
A_1 &= \frac{2}{d^2}\, (d+2) -\left( \frac{1}{2}\, \bcg_2(1) +\frac{ 2}{d}\, \bcv_2(1)\right) ,\\
A_2 &= -\frac{1}{2\,d} \ , \qquad B_1 = -2 \, A_3 = \frac{2}{d\, (d-2)} \ , \qquad B_2 = \frac{1}{d-2}  \ .
\end{split}
\label{}
\end{equation}	
Furthermore, it is possible to show that one can write the divergence of this gravitational entropy current as 
\begin{equation}
\WD_\mu J^\mu_S = \frac{2\,\eta}{T} \, \left(\sigma_{\mu\nu} + \frac{1}{2}\, \left[ \frac{d}{4\pi\,T}\, (1+ A_1 \,  d) - \tau_\pi\right]\, u^\alpha\, \WD_\alpha\sigma_{\mu\nu}\right)^2 + \cdots  \ ,
\label{divec2}
\end{equation}	
which is accurate up  to third order in the derivative expansion and clearly satisfies the requirement of non-negative divergence to that order.  We have recast this expression purely in terms of fluid dynamical variables and it generalizes the result for viscous relativistic fluids \req{divec1}.

\paragraph{Event horizon vs. quasi-local horizons:} We now turn  briefly to the subtlety mentioned at the beginning of this sub-section with regard to using the area of the event horizon for capturing the entropy of the dual field theory in situations out of global equilibrium. A key feature of the event horizon is its teleological nature, i.e., the fact that one needs to know the entire future evolution of the spacetime in order to determine its location. This aspect of event horizons generically implies that one typically may encounter a horizon in the bulk even before we perturb the system -- the event horizon grows in anticipation of the perturbation one is about to impart. Associating an entropy current with the event horizon as a result leads to a non-local and acausal definition of entropy as has been recently noted in \citep{Chesler:2008hg} (cf., \cite{Figueras:2009iu}). From an underlying statistical description, we would like however to define an entropy current a la Boltzmann, a local H-function which is defined purely based on the local dynamics of the fluid, with no recourse to late time boundary conditions. Moreover, there is a simple hydrodynamic flow on $\R^{d-1,1}$, the conformal soliton flow \citep{Friess:2006kw}, where it has been shown that the event horizon area does not capture the entropy of the dual field theory \citep{Figueras:2009iu}. Based on the latter analysis, it appears that in in certain dynamical situations one should use the area of apparent horizons (more precisely dynamical horizons), to define the entropy current.\footnote{Another piece of evidence in favour of quasi-local horizons comes from the analysis of holographic entanglement entropy  (for specified regions on the boundary) in time dependent states of the dual field theory, see the discussion in  \citep{Hubeny:2007xt}.} However, this statement glosses over certain subtleties involved in defining such quasi-local horizons\footnote{As explained in \citep{Figueras:2009iu} we use the term apparent horizon to denote a co-dimension one surface in the spacetime, which in conventional general relativity literature would correspond to a marginally outer trapped tube.} -- for an account of the issues involved we refer the reader to \citep{Figueras:2009iu}.

\section{Generalizations of the fluid-gravity correspondence}
\label{s:generalz}

In \sec{s:gravity} and \sec{s:grav2} we have discussed how one can construct gravitational solutions dual to fluid dynamics concentrating just on energy-momentum flow.  One of the advantages of focussing on the stress tensor $T^{\mu\nu}$ is that 
in the holographic context one deals purely with gravitational dynamics in \AdS{d+1}. This allows for an universal description; the Lagrangian \req{ehact} is a consistent truncation of the bulk string theory in the AdS/CFT context (all matter field interact at best quadratically with the gravitational field).  Nevertheless, one can incorporate other fields into the fluid-gravity correspondence and by now there is a large body of literature exploring these issues as discussed at the end of \sec{s:intro}.  We will describe the key features of these investigations using an abstract model and suggest some future generalizations.

\subsection{Fluid-gravity and the inclusion of matter}
\label{s:abstract}

Consider a situation where we have a gravitational action which is described by the Einstein-Hilbert action with negative cosmological constant coupled to some matter fields which we collectively denote as $\Psi$. To be precise,
\begin{equation}
\CS_{\text{bulk}} = \frac{1}{16\pi\,G_N^{(d+1)}}\,\int\, d^{d+1}x\, \sqrt{-G} \,\left(R - 2\,\Lambda\right)  + \CL_{\text{matter}}\left(\Psi, G_{MN}\right)
\label{ehactm}
\end{equation}	
where we have included metric contributions into the matter Lagrangian explicitly to allow for situations where we consider higher derivative theories of gravity (say for example the Gauss-Bonnet term in $d\ge 4$ as discussed in \citep{Dutta:2008gf}). 

We will assume that the action \req{ehactm} admits stationary asymptotically \AdS{d+1} black hole solutions\footnote{The general scheme we describe below is not restricted to asymptotically AdS spacetimes. The only reason to focus on these cases is the gauge-gravity duality which allows us to  relate the bulk gravity dynamics to that of a boundary field theory and in particular allows us via the holographic renormalization scheme to extract a boundary stress tensor. One can construct inhomogeneous dynamical black holes using the slow variation ansatz for any desired asymptopia (including asymptotically flat spacetimes). We will discuss an example with different asymptopia in \sec{s:nonrel}.} which correspond to global thermal equilibrium, which by picking suitable coordinates we can write as\footnote{In the following discussion we eschew the use of the Weyl covariant form since our disucssion would also be applicable to non-conformal fluids.}
\begin{equation}
\begin{aligned}
ds^2 &= -2 \, \CS(r, Q)\, u_\mu \, dx^\mu \, dr - r^2 \, \CV(r,Q)\, u_\mu \, u_\nu \, dx^\mu \, dx^\nu + r^2\, P_{\mu \nu}\, dx^\mu\, dx^\nu \\
\Psi & = \Psi(r,u^\mu, Q)
\end{aligned}
\label{genbga}
\end{equation}	
where we have once again resorted to the ingoing Eddington-Finkelstein coordinates. 
The coordinate $x^\mu$ as before span the boundary directions with $\mu = \{ 0, \cdots , d-1\}$.
Furthermore, we  assume that we have translational invariance in the spatial directions which we have exploited to boost the black hole solution so as  to allow for the velocity field. This of course requires that the matter field $\Psi$ supporting the black hole to respect the symmetries of the metric. We have captured all the parameters describing the black hole into a single variable $Q$, which enters into the metric and the matter field and corresponds to all the physical charges we wish to ascribe to the geometry, such as temperature $T$, Maxwell charges $q_I$ (or equivalently chemical potentials $\mu_I$), etc..
Depending on the boundary conditions on the matter field $\Psi$ we would find ourselves working in the canonical or the grand canonical ensemble.\footnote{Recall that for charged black holes which corresponds to the situation where $\Psi$ is a bulk Maxwell field, the grand canonical ensemble corresponds to non-zero value for the scalar potential $A_0$ which is the boundary chemical potential $\mu$.} It is possible to relax some of the assumptions described above at the expense of complicating the discussion. 

To obtain a geometry dual to fluid dynamics we want to promote the parameters of the background to fields as in conventional collective coordinate quantization. In order to do so we need to identify the full set of zero-modes. It is worthwhile to pause to take stock of some of the examples that have been discussed in the literature to orient ourselves:
\begin{enumerate}
\item In the forced fluid dynamics discussed in \citep{Bhattacharyya:2008ji}, $\Psi$ is just the dilaton and doesn't introduce new conserved quantities. So the parameter $Q$ is just the temperature of the \SAdS{} black hole.
\item For the $U(1)_R$-charged fluids discussed by \citep{Erdmenger:2008rm,Banerjee:2008th,Hur:2008tq} one has $\Psi$ to be the bulk Maxwell field which gives rise to an extra parameter, $q$, corresponding to the black hole charge. 
\item The recent analysis of \citep{Torabian:2009qk} incorporates three $U(1)$ R-charges in \AdS{5}. The matter content of the gravitation theory comprises of three Maxwell fields and three bulk scalars.  The black holes are described by four parameters, three charges and a temperature, collected here into $Q$ in addition to the velocity field.\footnote{Unlike the uncharged case discussed in \sec{s:grav2}, for the fluid solutions dual to gravity which have been constructed so far the issue of regularity is not clear; all the analyses so far do not demand regularity of the Maxwell potential on the future horizon.}
\item  One can also consider more exotic situations involving phase transitions. For instance in the Abelian-Higgs model coupled to gravity in \AdS{4} one has novel scalar hair black hole solutions \citep{Hartnoll:2008kx}. Here $\Psi$ comprises of bulk Maxwell field and a charged scalar field. The set of parameters $Q$ depends on which sector of the theory one considers since the bulk description involves non-trivial phase structure. In the $U(1)$ symmetric phase $Q $ just comprises of the temperature and chemical potential, whilst in the phase with broken $U(1)$, one  also has to incorporate the Goldstone mode. The latter phase gives rise to superfluid dynamics with a Landau two-fluid description. This has been discussed at linearized level in \citep{Herzog:2008he,Yarom:2009uq} (see also \cite{Basu:2008st}) and can indeed be incorporated into the fluid-gravity correspondence \citep{Hubeny:2009sf} (albeit with some effort since the background solutions are only known numerically).
\end{enumerate}

Having identified the set of parameters $Q$ for a given example, we are in a position to promote them to fields depending on the coordinates $x$ and proceed to solve the equations of motion \req{ehactm} order  by order in a derivative expansion. As described extensively above, this involves identifying two pieces of information. The first is to determine the linear operator ${\mathbb H}$ that acts on the correction terms to the metric. This operator by virtue of the Killing symmetries in the background spacetime (which we assume) and matter fields \req{genbga} is just a second order ordinary differential operator involving $r$. The second part involves identifying the source $s_n$ which involves a bit of  work at each order.  Schematically, if we promote  $Q \to Q(x)$ and $u_\mu \to u_\mu(x)$ and work by expanding the solution \req{genbga} in the variations about the background value. Specifically, consider
\begin{equation}
\begin{aligned}
u_\mu &= u^{(0)}_\mu + \delta u_\mu (x) =  u^{(0)}_\mu + \varepsilon\, x^\nu \p_\nu u_\mu (x)+\ord{\varepsilon^2}\ , \\
Q &= Q^{(0)} + \delta Q =\partial_\nu Q (x)  +\varepsilon\, x^\mu\, \partial_\mu Q^{(0)}  +\ord{\varepsilon^2}\ .
\end{aligned}
\label{}
\end{equation}	
For simplicity of discussion consider backgrounds which are stationary, so that by passing to the local inertial frame we can always choose $u_\mu^{(0)} = -\delta_\mu^v$ and $u^{(1)}_i = x^\mu\, \partial_\mu \beta_i$ as before using \req{defun}. Plugging these expansions into the metric we obtain the leading terms that contribute to the sources.
Using 
\begin{equation}
\CS(r, Q(x)) = \CS(r, Q^{(0)}) + \frac{\partial \CS(r,Q)}{\partial Q}\bigg|_{Q = Q^{(0)}} \, \delta Q 
\label{}
\end{equation}	
we obtain the leading order expansion of the background metric and matter fields
\begin{eqnarray}
ds^2 &=& 2\, \CS(r,Q)\, dv\,dr - r^2\, \CV(r,Q)\, dv^2 + r^2\, \delta_{ij}\, dx^i\,dx^j  \nonumber \\
&& +2\, \frac{\partial \CS}{\partial Q} \,\delta Q\, dv \,dr - 2\, \CS \, \delta \beta_i \,dx^i\,dr   -r^2\, \frac{\partial \CV}{\partial Q} \,\delta Q\, dv^2  -2 \,r^2\, \delta\beta_i \,(1-\CV) dx^i\,dv \nonumber \\
\Psi &=& \Psi(r,Q^{(0)}) + \frac{\partial \Psi}{\partial u^\alpha}\, \delta u^\alpha+ \frac{\partial \Psi}{\partial Q}\, \delta Q
\label{firstmet}
\end{eqnarray}	
which can then be used to solve for the higher order corrections to the metric $G^{(k)}$ and matter fields $\Psi^{(k)}$.  This has been carried out for a wide class of models and we refer the reader to the original literature described at the end of \sec{s:intro} for details of the construction in the specific cases. 

\paragraph{Universality of transport coefficients:} One of the interesting results arising from the explorations so far concerns the universality of transport coefficients in gravity duals. As is well known for two derivative gravity theories dual to boundary fluid dynamics, one has the famous ratio \citep{Kovtun:2004de} of shear viscosity to entropy density saturating the conjectured bound $\eta/s \ge 1/4\pi$. This can in fact be derived directly from the abstract discussion above, for the backgrounds \req{genbga} which respect spatial translational symmetry in global equilibrium for asymptotically \AdS{d+1} spacetimes  \citep{Haack:2008xx,Hubeny:2009sf}. 

It is interesting to ask whether the higher order transport coefficients exhibit similar properties. Based on the general analysis of gravity coupled to scalars and Maxwell fields and assuming that the theory doesn't exhibit any phase transitions (such as spontaneously Higgsing of the bulk gauge field), it was shown in \citep{Haack:2008xx} that 
\begin{equation}
4\, \lambda_1 + \lambda_2 = 2 \, \eta \, \tau_\pi
\label{}
\end{equation}	
which is of course satisfied by \req{transpgend}. 

\subsection{Non-relativistic fluids from gravity}
\label{s:nonrel}
We have concentrated thus far on the dynamics of relativistic conformal fluids and constructed holographic duals for them. We now briefly comment on deriving the dynamics of non-relativistic fluids from gravitational description. There are essentially two approaches taken in the literature so far: in \citep{Bhattacharyya:2008kq} gravity duals were obtained by taking a the non-relativistic limit of the relativistic system, while in \citep{Rangamani:2008gi} conformal non-relativistic fluids were discussed in the context of Galilean holography.

Non-relativistic ideal hydrodynamics is described by the continuity equation, 
\begin{equation}
\partial_t \rho_{\text{nr}} + \partial_i \left( \rho_{\text{nr}}\,  \vel^i\right) = 0,
\label{continuity}
\end{equation}
together with the equation of momentum conservation, (here $i = 1,\ldots,d$)
\begin{equation}
\partial_t (\rho_{\text{nr}}\, \vel^i ) + \partial_j \Pi^{ij} =0, \qquad
 \Pi^{ij} = \rho_{\text{nr}} \, \vel^i \, \vel^j + \delta^{ij} \, P_{\text{nr}}\,,
 \label{NS}
\end{equation}
and the equation of energy conservation,
\begin{equation}
  \partial_t \left( \epsilon_{\text{nr}} + \frac{1}{2}\, \rho_{\text{nr}} \, \vsq \right) + \partial_i \, j_\epsilon^i =0,
  \qquad
j_\epsilon^i =\frac{1}{2}\, ( \epsilon_{\text{nr}}+ P_{\text{nr}} )\, \vsq\, \vel^i \ .
\end{equation}
where $\vsq = \vel^i\, \vel^i$. Here $\rho_\text{nr}$ is the particle number density and $P_{\text{nr}}$, $\epsilon_{\text{nr}}$ is the pressure and energy density of the non-relativistic system under consideration and we use $v^i$ to  denote the non-relativistic velocity. These equations can be derived from the relativistic stress tensor \req{ideal1} by writing the conservation equation \req{cons1} in light-cone coordinates and demanding that the fluid variables be independent of one of the light-cone directions. Using the light-cone version of \req{cons1}
\begin{equation}
 \p_+ T^{++} + \p_i T^{+i} =0 \ , \qquad 
 \p_+ T^{+i} + \p_j T^{ij} = 0 \ , \qquad 
 \p_+ T^{+-} + \p_i T^{-i} =  0,
\label{releqns}
\end{equation}	
we can map the relativistic system in $d$ dimensions into the non-relativistic equations in $d-2$ spatial directions with the following identifications: $T^{++}$ is identified with the mass density, $T^{+i}$ with the mass flux (which is equal to the momentum density), $T^{ij}$ with the stress tensor, $T^{+-}$ with the energy density, and $T^{-i}$ with
the energy flux,
\begin{eqnarray}
  && T^{++} =\rho_\text{nr}, \quad T^{+i} = \rho_\text{nr} \, \vel^i, \quad T^{ij} =\Pi^{ij} ,
  \nonumber \\
  && T^{+-} = \epsilon_\text{nr} + \frac{1}{2}\, \rho_\text{nr} \, \vsq, \qquad T^{-i} = j_\epsilon^i .
  \label{stiden}
\end{eqnarray}

It is now easy to convince oneself based on \req{stiden} that the
precise mapping between relativistic and non-relativistic hydrodynamic
variables is
\begin{eqnarray}
  \u^+ & = &\sqrt{ \frac{1}{2} \, \frac{\rho_\text{nr}}{ \epsilon_\text{nr} + P_\text{nr}} } \ , \qquad 
  \u^i  = \u^+ \,\vel^i ,\nonumber\\
  P &=& P_\text{nr} \, , \qquad \qquad\quad\;\;
  \rho = 2\,\epsilon_\text{nr} + P_\text{nr}.
\label{idealmap}
\end{eqnarray}
The component of the relativistic velocity $\u^-$ can be determined
using the normalization condition $\u_\mu\,\u^\mu = -1$ to be
\begin{equation}
 \u^- = \frac{1}{2}\, \left( \frac{1}{\u^+} + \u^+ \,\vsq \right) .
\label{ummap}
\end{equation}	

This maps makes it clear that the transport coefficients of the non-relativistic theory are inherited from the parent relativistic hydrodynamics. In this description it is clear that non-relativistic fluids with holographic duals will saturate the conjectured viscosity bound $\eta/s = 1/4\pi$, which was verified explicitly in \cite{Herzog:2008wg,Adams:2008wt}. Furthermore, it is also possible to use this light-cone reduction to infer the heat conductivity of the non-relativistic fluid:
\begin{equation}
\kappa_\text{nr} = 2\,\eta_\text{nr}\, \frac{\epsilon_\text{nr}+P_\text{nr}}{\rho_\text{nr}\,  T}.
\label{kappaval}
\end{equation}
which can be rephased as the statement that the Prandtl number of the fluid is unity. We recall that the Prandtl number is defined as the ratio of the kinematic viscosity $\nu_\text{nr}$ and the thermal diffusivity $\chi_\text{nr}$,
\begin{equation}
  {\rm Pr} = \frac{\nu_\text{nr}}{\chi_\text{nr}},
 \label{Prdef}
\end{equation}
where
\begin{equation}
  \nu_\text{nr} = \frac{\eta_\text{nr}}{\rho_\text{nr}}\ , \qquad \chi_\text{nr} = \frac{\kappa_\text{nr}}{\rho_\text{nr} \, c_p},
\end{equation} 
where $c_p$ is the specific heat at constant   pressure. 

The general idea of using slow variation in certain directions to construct inhomogeneous dynamical black hole spacetimes is of course not restricted to asymptotically AdS spacetimes and one can exploit this scheme to construct new solutions with different asymptotics.  For general asymptotics one does not recover any interesting boundary dynamics. However, in recent years we have seen interesting generalizations of the AdS/CFT correspondence, such as the holographic duals for field theories with non-relativistic Schr\"odinger symmetry as originally discussed in \citep{Son:2008ye,Balasubramanian:2008dm} and later embedded into string theory in 
\citep{Herzog:2008wg, Maldacena:2008wh,Adams:2008wt}. In these examples one has spacetimes with non-trivial asymptotic structure and the dual field theory is a non-local deformation of a relativistic superconformal field theory such as $\CN=4$ SYM. 

In \citep{Rangamani:2008gi} the holographic duals for fluids with non-relativistic conformal symmetry i.e., Schr\"odinger invariance was constructed. 
In this case one can achieve the holographic dual in two equivalent ways: either by implementing the general procedure outlined in the lectures (taking into account the reduced symmetries of the problem) or equivalently by exploiting a specific duality transformation in string theory, the so called Null Melvin Twist \citep{Gimon:2003xk} or the TsT transformation \citep{Maldacena:2008wh}. To be specific, \citep{Rangamani:2008gi} constructed fluid dynamical solutions for the five-dimensional effective action involving a scalar field $\phi$ and a massive vector $A_M$
\begin{equation}
  S = \frac{1}{16 \, \pi G_N^{(5)}}\,\int d^5 x \sqrt{-G} \left(R 
  - \frac{4}{3} (\partial_M \phi) (\partial^M \phi) 
  - \frac{1}{4} e^{-8 \phi / 3} F_{MN} F^{MN} 
- 4 \,A_M A^M - V(\phi) \right)  ,
\label{5deff}
\end{equation}
where
\begin{equation}
V(\phi) = 4 \,e^{2 \phi/3} (e^{2 \phi} - 4) \ , 
\label{Vpot}
\end{equation}	
which asymptote to the \Schr{5} spacetime,
\begin{equation}
ds^2 = r^2\, \left(-2 \, dx^+\, dx^- - \beta^2 \,r^{2}\, (dx^+)^2 
+ d{\bf x}_2^2 \right) + \frac{dr^2}{r^2}.
\label{vacmet}
\end{equation}	

Consider the general form of the fluid dynamical metrics in \AdS{5} given by \req{formmet} with $d=4$. We introduce light cone coordinates $x^\pm$ on the boundary directions.   Restricting to configurations which have
$\partial_-$ as an isometry, after a TsT transformation on the metric \req{formmet}
we get a new metric of the form 
\begin{eqnarray}
  ds_E^2 &=& e^{-\frac{2}{3} \, \phi} \, \left( 
  -2\, \u_\mu\, \CS\, dx^\mu \, dr + \left[\chi_{AB} 
  - \frac{{\widetilde\chi}_A\, {\widetilde \chi}_B}
  {1+ \chi_{--}}  \right]\, dX^A\, dX^B \right), \nonumber \\
    A &=&  e^{2\phi}\, \widetilde{\chi}_A\, dX^A, \nonumber \\
  e^{2\,\phi} &=& \frac{1}{1+ \chi_{--}} ,
\label{nrflumet}
\end{eqnarray}	
with 
\begin{equation}
{\widetilde \chi}_A = \chi_{A-} - \u_- \, \CS \, \delta_A^r \ .
\label{tchidef}
\end{equation}	
The TsT transform converts the asymptotically \AdS{5} spacetime
\req{formmet} to an asymptotically \Schr{5} spacetime, which depends
on the paramters $b, \beta, \u_i$ which are arbitrary functions of $(x^+, {\bf x})$. 
The relativistic velocity field $u^\mu$ in fact descends naturally into a non-relativistic velocity field $\vel_i$ via the map
\begin{equation} \label{bulkmapp}
\u^+ = \beta, \quad \u^i =  \beta \left[ \vel^i +
  \frac{1}{4 \, \beta^2} \, \partial_i (b\,\beta)\right]. 
\end{equation}
which can be inferred from  the light-cone reduction of the stress tensor \citep{Rangamani:2008gi}. In the field theory description $b$ is related to the temperture, while $\beta$ sets the chemical potential for the non-relativistic particle number.

Conformal non-relativistic fluids with Schr\"odinger symmetry are in general compressible; this follows from the fact that the energy density is related to the pressure $2 \, \epsilon_{\text{nr}} = d\, P_\text{nr}$ through the equation of state (which in turn follows from scale invariance). To make contact with the usual studies of incompressible
Navier-Stokes equations we need to ensure that we can decouple the
fluctuations in the density. This can be achieved by looking at low
frequency modes which do not excite the propagating sound mode in a
hydrodynamic system, i.e., by focussing on the shear mode.  In fact, this limit was discussed recently in the context of the fluid-gravity correspondence in
\citep{Bhattacharyya:2008kq} (see also \citep{Fouxon:2008tb,Fouxon:2008ik} for closely related results), where the authors showed
that starting from a parent relativistic conformal fluid dynamical
system one can recover incompressible Navier-Stokes equations in a
suitable scaling limit. This scaling limit reveals an interesting structure in the fluid equations -- they are scale invariant under a new scaling symmetry, which is the Galilean conformal algebra discussed recently in \citep{Bagchi:2009my}. This symmetry is different from the Schr\"odinger symmetry enjoyed by the conformal non-relativistic  fluids discussed above (see also \cite{Gusyatnikova:1989nx}, \cite{Hassine:1999ab}).

One can view these two constructions as follows: given a relativistic theory, in particular relativistic fluid dynamics, one can attain a non-relativistic limit either by (i) taking the speed of light to infinity or (ii) by a light-cone reduction. The former procedure is related to contracting the Poincar\'e algebra to the Galilean algebra and when implemented on the relativistic Navier-Stokes equations leads to the incompressible non-relativistic Navier-Stokes fluid of \citep{Bhattacharyya:2008kq}. The latter procedure of light-cone reduction converts $d$ dimensional relativistic fluid dynamics into a $d-2$ spatial dimensional non-relativistic fluid dynamics. Since one requires a null vector to reduce on the light-cone, we end up losing an effective dimension in our field theoretic description. 

\section{Discussion}
\label{s:discuss}

In the course of these lectures we have discussed the essential features which relate the physics  of inhomogeneous, dynamical, black hole solutions of general relativity in asymptotically AdS spacetimes, to fluid dynamics of strongly coupled superconformal field theories living on the boundary of these AdS geometries. In particular, given any solution to the relativistic Navier-Stokes equations and their non-linear generalizations (with the strong coupling values of the transport coefficients),   one can construct an asymptotically AdS black hole solution. Alternately, the general construction described here can be viewed as a derivation of the hydrodynamic stress tensor of the dual superconformal field theories. While we have discussed the construction to second order in derivatives, it is clear that the general  construction can in principle be extended to arbitrary orders in the gradient expansion (albeit with increasing computational complexity in evaluating the source terms). 

We have also described how one can use the geometric description to understand the field theory entropy and constructed a specific gravitational entropy current. 
An  important issue which we have briefly discussed concerns the geometric description of entropy and subtleties associated with identifying the entropy with the area of the event horizon. While it seems reasonable to associate  the area of the event horizon to the field theory entropy in situations where one has slow variations, it seems clear that this cannot be true in general as exemplified by the conformal soliton flow described in \cite{Figueras:2009iu}.

Furthermore, we have reviewed various generalizations of the correspondence over the past year or so which have led to interesting new insights into forced fluid dynamics and charge transport, etc.. These provide an interesting arena for further exploration -- in principle it should be possible to derive the complete stress tensor and charge currents for $\CN=4$ SYM up to two derivatives incorporating the three $U(1)$ R-charge and angular momentum chemical potential.  In addition, it is clear that there are many interesting directions that can be tackled within this framework, most notably the issues which were raised in \sec{s:intro} as part of the motivation for the correspondence.  

One issue we have not touched upon in these lectures is the relation of the fluid-gravity correspondence to the membrane paradigm \cite{Thorne:1986eu}. Both purport to identify the dynamics of black holes with hydrodynamics, and indeed the correspondence offers a important new perspective on this issue \cite{Hubeny:2009zz}.\footnote{An earlier discussion of the membrane paradigm in the context of hydrodynamics in AdS/CFT can be found in  \cite{Kovtun:2003wp}. Also see \cite{Eling:2009oz} for a recent discussion of  the membrane paradigm in the holographic context.} Since in the fluid-gravity correspondence the entire spacetime dynamics is mapped unambiguously into the boundary fluid dynamics, it is natural to argue that the correspondence in fact implements the ideas inherent in the membrane paradigm, albeit with a new wrinkle: the membrane is not located at the stretched horizon, but rather at the boundary of the spacetime.  Despite its location, this `membrane at the end of the universe' faithfully captures the entire bulk spacetime dynamics  and implements the membrane paradigm holographically, as is clear from the fact that we not only recover the stress tensor of the fluid, but also a particular gravitational entropy current. Note that in the conventional membrane paradigm, the dynamics of the stretched horizon only captures the physics of the region behind the black hole horizon and not of the entire spacetime. One can of course use the AdS/CFT correspondence to interpolate between these two extreme descriptions \cite{Iqbal:2008by}: if one so wishes, it is possible to define a fiducial membrane at some other radial location (and in particular on the stretched horizon) and identify the fluid dynamics on this surface. The dynamics on this surface is related to the boundary fluid dynamics by an appropriate flow equation which can be derived from the bulk gravitation equations using techniques analogous to the holographic renormalization group.

Finally, we should mention recent investigations of gravitational duals of field theories perturbed away from equilibrium \citep{Chesler:2008hg, Bhattacharyya:2009uu}. As we have emphasized the hydrodynamic behaviour comes into play only when the field theory achieves local thermal equilibrium. In general, large perturbations will evolve outside the fluid regime, till such a time the system thermalizes and achieves local equilibrium. 
Correspondingly, in the dual gravity description,  black hole formation is outside the long-wavelength hydrodynamic regime. The remarkable aspect of the recent discussion of \cite{Bhattacharyya:2009uu} is that the system tends to ``thermalize'' rapidly -- in effect the hydrodynamic description takes over almost instantaneously after the perturbation is switched off.\footnote{To be sure, this is a statement that the one point functions display thermal behaviour at time-scales much shorter than that set by the thermal wavelength. Furthermore, this statement is seen to hold only for small amplitudes of the perturbation.} It would be interesting to understand this passage to local thermal equilibrium for a generic perturbation in a strongly coupled field theory in greater detail.

\subsection*{Acknowledgements}

It is a pleasure to thank my collaborators, Sayantani Bhattacharyya, Veronika Hubeny, R. Loganayagam, Gautum Mandal, Shiraz Minwalla, Takeshi Morita,  Harvey Reall, Simon Ross, Dam Son,  Ethan Thompson and Mark Van Raamsdonk for wonderful collaborations on various aspects of fluid dynamics. I would also like to thank the students of the CERN Winter school for their enthusiasm and detailed questions. It is furthermore a great pleasure to the organizers, especially Angel Uranga for putting together an excellent winter school. I would like to thank CERN and KITP for their hospitality. This work is supported in part by STFC Rolling grant and by the National Science Foundation under the Grant No. NSF PHY05-51164.

\appendix
\section{Weyl covariant curvature tensors}
\label{a:wcurve}
One can define a curvature associated with the Weyl covariant derivative by the usual procedure of evaluating the commutator between two covariant derivatives. 
Defining a field strength for the Weyl connection
\begin{equation}
\CF_{\mu \nu}  = \nabla_\mu \CA_\nu - \nabla_\nu \CA_\mu
\label{wfield}
\end{equation}	
we find a Riemann curvature tensor $\CR_{\mu\nu\lambda\sigma}$:
\begin{equation}
\CR_{\mu\nu\lambda\sigma}= R_{\mu\nu\lambda\sigma} +4\, \delta^\alpha_{[\mu}g_{\nu][\lambda}\delta^\beta_{\sigma]}\left(\nabla_\alpha \CA_\beta + \CA_\alpha \CA_\beta - \frac{\CA^2}{2} g_{\alpha\beta} \right) - \CF_{\mu\nu}\, g_{\lambda\sigma}
\label{wriemann}
\end{equation}	
These two tensors are in an appropriate sense Weyl invariant; it is not hard to check from the definitions that  $ \CF_{\mu\nu}=\widetilde{\CF}_{\mu\nu}$ and $\CR_{\mu\nu\lambda}{}^\alpha=\widetilde{\CR}_{\mu\nu\lambda}{}^\alpha$. 
It should be borne in mind that the Weyl covariant Riemann tensor defined above \req{wriemann} has different symmetry properties from the conventional Riemann tensor. Most importantly, it is not anti-symmetric under the exchange of the last two indices. For example, 
\begin{equation}\label{curvsym:eq}
\begin{split}
\CR_{\mu\nu\lambda\sigma}+\CR_{\mu\nu\sigma\lambda} &= -2\, \CF_{\mu\nu} \,g_{\lambda\sigma} \\
\CR_{\mu\nu\lambda\sigma}-\CR_{\lambda\sigma\mu\nu} &= \delta^\alpha_{[\mu}g_{\nu][\lambda}\delta^\beta_{\sigma]} \,\CF_{\alpha\beta} - \CF_{\mu\nu}\, g_{\lambda\sigma} + \CF_{\lambda\sigma} \,g_{\mu\nu} \\
\CR_{\mu\alpha\nu\beta}\,V^\alpha \,V^\beta-\CR_{\nu\alpha\mu\beta}\,V^\alpha \,V^\beta &= - \CF_{\mu\nu}\ V^\alpha \,V_\alpha\\
\end{split}
\end{equation}

With the basic definitions \req{wfield} and \req{wriemann} in place we can proceed to define analogous expressions for various other tensors often encountered in differential geometry, such as  the  Ricci tensor, Ricci scalar:
\begin{equation}
\begin{split}
\CR_{\mu\nu}&\equiv \CR_{\mu\alpha\nu}{}^{\alpha} = R_{\mu\nu} -(d-2)\left(\nabla_\mu \CA_\nu + \CA_\mu \CA_\nu -\CA^2 g_{\mu\nu}  \right)-g_{\mu\nu}\nabla_\lambda\CA^\lambda - \CF_{\mu\nu} = \widetilde{\CR}_{\mu\nu}\\
\CR &\equiv \CR_{\alpha}{}^{\alpha} = R -2(d-1)\nabla_\lambda\CA^\lambda + (d-2)(d-1) \CA^2 = e^{-2\phi} \widetilde{\CR}\ .\\
\end{split}
\label{ricEin:eq}
\end{equation}
In addition it is  also worth noting that the conformal tensors of the underlying spacetime manifold appear as a subset of conformal observables in hydrodynamics. These conformal tensors are the Weyl-covariant tensors that are independent of the background fluid velocity, for we have already accounted for the terms involving velocity derivatives explicity above. A well known example of this class of operators is the Weyl curvature $C_{\mu\nu\lambda\sigma}$ which is the trace free part of the Riemann tensor. In $d\geq3$ it is given as 
\begin{equation}\label{weylcurv:eq}
\begin{split}
\CC_{\mu\nu\lambda\sigma} &\equiv \CR_{\mu\nu\lambda\sigma}+4\,\delta^\alpha_{[\mu}g_{\nu][\lambda}\delta^\beta_{\sigma]}\mathcal{S}_{\alpha\beta} = C_{\mu\nu\lambda\sigma} - \CF_{\mu\nu} g_{\lambda\sigma} = e^{2\phi}\,\widetilde{\CC}_{\mu\nu\lambda\sigma}
\end{split}
\end{equation}
where we introduced the Schouten tensor $\CS_{\mu\nu}$ is defined as\footnotemark
\begin{equation}\label{schouten:eq}
\begin{split}
\CS_{\mu\nu} &\equiv \frac{1}{d-2}\left(\CR_{\mu\nu}-\frac{\CR g_{\mu\nu}}{2(d-1)}\right) = S_{\mu\nu}-\left(\nabla_\mu \CA_\nu + \CA_\mu \CA_\nu - \frac{\CA^2}{2} g_{\mu\nu} \right) -\frac{\CF_{\mu\nu}}{d-2} =\widetilde{\mathcal{S}}_{\mu\nu}
\end{split}
\end{equation}

\footnotetext{Often in the study of conformal tensors, it is useful to rewrite other curvature tensors in terms of the Schouten and the Weyl curvature tensors-
\begin{equation}\label{weylschouten:eq}
\begin{split}
\CR_{\mu\nu\lambda\sigma}&=\mathcal{C}_{\mu\nu\lambda\sigma} -\delta^\alpha_{[\mu}g_{\nu][\lambda}\delta^\beta_{\sigma]}\mathcal{S}_{\alpha\beta},\qquad \CR= 2(d-1)\mathcal{S}_\lambda{}^\lambda \\
\CR_{\mu\nu}&= (d-2) \mathcal{S}_{\mu\nu} + \mathcal{S}_\lambda{}^\lambda g_{\mu\nu} ,\qquad  \mathcal{G}_{\mu\nu}= (d-2)(\mathcal{S}_{\mu\nu}-\mathcal{S}_\lambda{}^\lambda g_{\mu\nu}) \\
\end{split}
\end{equation}}

From equation \eqref{weylcurv:eq}, it is clear that $ C_{\mu\nu\lambda\sigma}=\mathcal{C}_{\mu\nu\lambda\sigma} +\CF_{\mu\nu} g_{\lambda\sigma} $ is clearly a conformal tensor. Such an analysis can in principle be repeated for the other known conformal tensors in arbitrary dimensions.

The Weyl Tensor $ C_{\mu\nu\lambda\sigma}$ has the same symmetry properties as that of Riemann Tensor $ R_{\mu\nu\lambda\sigma}$,
\begin{equation}
\begin{split}
C_{\mu\nu\lambda\sigma} &= -C_{\nu\mu\lambda\sigma} = -C_{\mu\nu\sigma\lambda}=C_{\lambda\sigma\mu\nu} \\
\qquad &\text{and}\quad C_{\mu\alpha\lambda}{}^\alpha =0 \ . 
\end{split}
\label{wweyl}
\end{equation}
It follows that $C_{\mu\alpha\nu\beta}\,u^\alpha\, u^\beta$ is a symmetric traceless and transverse tensor, which is why it shows up as a second order contribution in the conformal hydrodynamical stress tensor.


\providecommand{\href}[2]{#2}\begingroup\raggedright\endgroup

\end{document}